\documentclass[11pt]{article}
\usepackage{comment, lscape}
\usepackage[nodisplayskipstretch]{setspace}
\usepackage{url}
\usepackage[hidelinks]{hyperref}
\usepackage{appendix}
\usepackage[active]{srcltx}

\usepackage{graphicx}
\usepackage{graphics}
\usepackage{caption}
\usepackage{subcaption}
\usepackage{rotating}
\usepackage{verbatim}
\usepackage{booktabs}
\usepackage{xltabular}
\usepackage{natbib}
\usepackage{supertabular}
\usepackage{bbm}
\usepackage{dsfont}
\usepackage{amsmath, amssymb, amstext,amsfonts}
\usepackage{xcolor}
\usepackage{soul}
\PassOptionsToPackage{hyphens}{url}\usepackage{hyperref}
\hypersetup{
	colorlinks=true,
    citecolor=blue,
    linkcolor=blue,
    urlcolor=blue
}

\usepackage{bibunits}

\defaultbibliography{Muni_adv_MFJ_R2}
\defaultbibliographystyle{chicago}

\usepackage{tikz}
\usetikzlibrary{calc}
\usepackage{titlesec}
\newcolumntype{H}{>{\setbox0=\hbox\bgroup}c<{\egroup}@{}}
\usepackage{pdflscape}

\usepackage[margin=1.0in]{geometry}

\newcommand{\RN}[1]{\uppercase\expandafter{\romannumeral #1\relax}}

\usepackage{setspace}

\usepackage{endnotes}

\makeatother
\begin{document}

\vspace{-24pt}
\title{
\LARGE \bf Fiduciary Duty in the Municipal Bonds Market
\thanks{\rm I am highly indebted to my advisor, Sudheer Chava, for his invaluable guidance, support, and encouragement. I am very grateful to my dissertation committee members, Rohan Ganduri, Manasa Gopal, Alexander Oettl, and Manpreet Singh, for their constant advice and help. I benefited from discussions and comments from Rajesh Aggarwal, Nicole Boyson (discusssant), Daniel Garrett,  Nikhil Paradkar, Ajay Subramanian, Daniel Weagley, and Linghang Zeng. I am thankful to conference participants at the FMA Doctoral Student Consortium 2022, and FMA Special PhD Paper Presentation 2022 for valuable comments. For insightful discussions on municipal bond issuance, I am indebted to Courtney Knight (Treasury, Debt and Investment Chief, City of Atlanta), Rachel Miller (Chief Deputy State Treasurer, New Hampshire State Treasury), Diana Pope (Director of Financing and Investment, Georgia State Financing and Investment Commission), Philip Potter (Georgia Municipal Association), and Bjorn Reigle (Managing Director, The PFM Group). I am also thankful to Jonathan Kugel and Arushi Saxena for their help in making the FOIA  requests in Virginia and Tennessee, respectively.}
}

	
	  \doublespacing

\vspace{-24pt}

\author{\small Baridhi Malakar \thanks{Scheller College of Business, Georgia Institute of Technology \textit{Email:} \texttt{baridhi.malakar@scheller.gatech.edu}}}

\date{}

\maketitle
\thispagestyle{empty}

\vspace{-24pt}
\begin{center}
\small First Draft: November 1, 2022\\
\small Current Draft: June 14, 2024       \\       
\smallskip

\end{center}



\begin{abstract}
	{\normalsize
		\onehalfspacing


\noindent I examine whether the imposition of fiduciary duty on municipal advisors affects bond yields and advising fees. Using a difference-in-differences analysis, I show that bond yields reduce by $\sim$9\% after the imposition of the SEC Municipal Advisor Rule due to lower underwriting spreads. Larger municipalities are more likely to recruit advisors after the rule is effective and experience a greater reduction in yields. However, smaller issuers do not experience a reduction in offering yields after the SEC Rule. Instead, their borrowing cost increases if their primary advisor exits the market. Using novel hand-collected data, I find that the average advising fees paid by issuers does not increase after the regulation. Overall, my results suggest that while fiduciary duty may mitigate the principal-agent problem between some issuers and advisors, there is heterogeneity among issuers.

\bigskip

}
\end{abstract}

\clearpage
\setcounter{page}{1}
\setcounter{secnumdepth}{3} \setcounter{tocdepth}{3}
\doublespacing

 \maketitle

%

\clearpage\newpage
\setcounter{page}{1}
\setcounter{secnumdepth}{3} \setcounter{tocdepth}{3}
\section{\label{sec:intro}Introduction}

State and local governments in the United States finance various infrastructure and public utility projects through municipal bonds. As of 2023, \$4 trillion of municipal bonds are outstanding, of which \$390 billion were issued in 2022. Municipal issuers may not always have the necessary in-house expertise \citep{doty2018regulation}. They typically hire advisors to help decide the method of sale (negotiated versus competitive \citep{marlowe2009method}), to structure the bonds and to develop and draft the offering statements \citep*{bergstresser2018evolving}. 

However, due to misaligned incentives, decentralized information, and missing internal controls \citep{garrett2021conflicts,park2017internal}, advisors may not always act in the interest of bond issuers\endnote{For instance, SEC Commissioner Kara M. Stein \href{https://tinyurl.com/2vws776j}{remarked in 2013}, that the SEC had observed ``numerous examples of bad behavior, including self-dealing and excessive fees''. Also see comments by  \href{https://tinyurl.com/yc6s4mxy}{Municipal Securities Rulemaking Board (MSRB)} and \href{https://www.bondbuyer.com/news/sec-taking-notice-of-unqualified-municipal-advisors}{others}.}. As a result of these potential principal-agent problems, the SEC imposed fiduciary duty on municipal advisors through the Municipal Advisor Rule (MA Rule) effective July 1, 2014. In this paper, I study how this imposition of fiduciary duty on the municipal advisors affects the offering yields and advising fees paid by the municipal issuers.






Using a canonical difference-in-differences research design, I compare the yield spreads for municipal bonds sold via negotiation to bonds sold via competitive bidding, before and after the SEC MA Rule. Bonds sold via negotiation are more likely to involve the underlying friction of misaligned incentives between the issuer and the advisor (treated group). This friction may be lower in bonds sold via competitive auction (control group), where price discovery happens without negotiation. I use this quasi-natural setting to understand the overall effect on bond yield spreads after the imposition of fiduciary duty on municipal advisors. 


The preferred specification indicates that offering yields decreased by 10.6 basis points (bps) for negotiated bonds relative to competitively bid bonds after the regulation. Given that municipalities may select into negotiating their bond sale, the baseline identification relies on within-issuer variation across advisors over time. Further, I absorb time-varying unobserved factors among advisors and among issuers based on the geographical state to which they belong. Finally, I include bond-level and other controls to account for observable characteristics. Considering the average bond in the sample, this effect amounts to a nearly 8.7\% reduction (=10.62/122) in yield spreads. 




The lower yields are driven by reduced underwriting spreads of negotiated bonds at the time of issuance. With advisors' obligation to adhere to their fiduciary duty, they are more likely to negotiate higher prices (lower yields), thereby reducing underwriting spreads. The effect is driven by issuers for whom advisors likely play a greater role in selecting underwriters. Large and sophisticated issuers seem to benefit more, while small issuers face higher borrowing costs due to the exit of advisors. Overall, offering yields reduce on average but there is heterogeneity among issuers.

\subsection{\label{subsec:lit_rev}Literature Review}

This paper contributes to three strands of the literature in finance and economics. First, this paper is related to the economics of expert advice in financial decision-making. \cite{inderst2012competition,inderst2012not} provide theoretical underpinnings to how competition and compensation structure are related to advice provided by financial intermediaries. Empirical work in this direction has shed light on advisors' commissions and other incentives in offering advice rather than clients' interests \citep{vijayakumar2006role,allen2010does,christoffersen2013consumers,anagol2017understanding,dimmock2018fraud,boyson2019worst,dimmock2021real}. This paper shows how enforcing discipline among municipal advisors through fiduciary duty may improve the average outcome for municipal borrowers. Importantly, I introduce novel data on financial intermediaries' municipal advising fees to show that advising fees does not increase after the MA Rule.


Second, this paper follows a recent surge in research on the municipal bond market \citep{cohen1989municipal,hildreth2002debt,hildreth2005evolution,cohen2013municipal,cohen2015holds,gaffney2016municipal,cohen2017headline,cohen2017rating,beatty2019rating,luby2019nic,bronshtein2020declining,azarmsa2021financing,mughan2021municipal,yang2021auditor,cms2022_bankruptcy,rizzi2022nature,neumann2022municipal,hazinski2022local,bruno2024environmental,chava2024impact}. Prior work leveraging the municipal bond market transactions has looked at liquidity \citep{harris2006secondary,schultz2012market,marlowe2013municipal,marlowe2020municipal} and default risk \citep{schwert2017municipal}, as well as tax-effects \citep{green1993simple,landoni2018tax,babina2021heterogeneous}. \cite{smull2023climate} show the relationship between physical climate risk and race on municipal credit risk. Research has also shown how local and fiscal conditions affect municipal bonds \citep{marlowe2007much,luby2012federal,yang2017financial,yang2020general,painter2020inconvenient,yang2022municipal}. In this context, this paper sheds new light on how federal regulation may reduce offering yields on average due to lower underwriting spreads. 

Finally, this paper contributes by showing the policy relevance of fiduciary duty. Recent work in this direction has estimated higher risk-adjusted returns by extending fiduciary duty \cite{egan2019brokers}. \cite{bhattacharya2019fiduciary} identify the effect of fiduciary duty in the reduced form by accounting for the entry margin using a structural model. Among municipal bonds, \cite{garrett2021conflicts} provides empirical evidence on how reducing conflicts of interest for advisors may improve bond outcomes. I contribute to this literature by showing how regulation may mitigate the principal-agent problem between issuers and advisors but has a differential effect among issuers. While large and sophisticated issuers experience lower offering yields after the fiduciary rule on municipal advisors, the same is not true for smaller issuers. 


\section{\label{sec:method}Theoretical Framework and Empirical Methodology}

There are competing views on the implications of imposing a fiduciary duty \citep{bhattacharya2019fiduciary}. Fiduciary duty may benefit issuers by alleviating the principal-agent problem and making it costly to offer poor advice (\emph{advice channel}). Advisors may be held liable for adverse consequences to issuers during bond issuance if they fail to adhere to their fiduciary responsibility after the MA Rule (see Section \ref{sec:sec_ma_rule} in the Internet Appendix for details). Alternatively, fiduciary duty may impose an undue burden on municipal advisors by increasing their costs (\emph{fixed costs channel}), leading to worse outcomes for the issuer. The overall effect on issuers from the imposition of fiduciary duty on municipal advisors is unclear.


I leverage a key feature of the municipal bond market to understand this overall effect. Municipal bonds are commonly sold via one of the two methods: competitive bidding or negotiated sale \citep{marlowe2009method}. Competitive bidding involves underwriters submitting their bids to buy the newly issued bonds. In a negotiated sale, a pre-selected underwriter works with the issuer to arrive at the terms of the sale. Municipal advisors help issuers in advertising the bond sale to potential underwriters in a competitive sale as well as in evaluating the final bids submitted. For negotiated sales, advisors may be involved from the time of selection of the underwriter to the final closure of the bond sale. 



The primary friction arises from the principal-agent problem between issuers and advisors during municipal bond issuance. When hiring the more-informed advisors, municipal issuers may not be able to monitor advisors. By requiring municipal advisors to owe a fiduciary duty to their clients, the SEC Municipal Advisor Rule addresses this friction by mitigating the principal-agent problem. This paper considers a quasi-natural experiment where the underlying information asymmetry between the issuer and the advisor is potentially lower in a competitive auction of bonds than in a negotiated sale. 




To formally characterize the baseline specification for the causal analysis of the Municipal Advisor Rule in a standard difference-in-differences \citep{bertrand2004much} equation, consider:
\begin{equation}\label{eq:baseline}
\begin{split}
y_{b,i,a,t} &= \alpha + \beta_0*Post_{t} \times Nego._{b,i,a} + \beta_1*Post_{t} + \beta_2*Nego._{b,i,a}  
 + X_{b} + Z_{i,t} \\  &+  \gamma_{i} + \mu_{a,t} + \kappa_{s,t} + \epsilon_{b,i,a,t}       
\end{split}    
\end{equation}
where index $b$ refers to bond, $i$ refers to issuer, $a$ denotes the municipal adviser and $t$ indicates time. The main outcome variable in $y_{b,i,a,t}$ is the offering yield spread at which a bond is issued in the primary market. $X_{b}$ includes control variables at the bond level that influence its value. These include the coupon (\%); log(amount issued in \$); dummies for callable bonds, bond insurance, general obligation bond, bank qualification, refunding, and credit enhancement; credit rating; remaining years to maturity; and inverse years to maturity (see Table \ref{table:vardescription} for detailed description). $Z_{i,t}$ corresponds to local economic conditions of the issuer's county/state. Following \cite{gao2019financing}, I use the lagged values for log(labor force) and unemployment rate, and the percentage change in unemployment rate and labor force, respectively\endnote{I aggregate county level metrics to arrive at the state level measures and use these for corresponding state level bonds.}. Additional controls take care of the general interest rate environment (captured by changes in the Federal Funds rate and 10-year US Treasury yields) as well as the quality of the underwriter (based on market share). 

This specification includes three sets of fixed effects. First, $\gamma_{i}$ indicates issuer fixed effects, to make comparisons within a given issuer. I also include fixed effects at the advisor $\times$ year ($\mu_{a-t}$) level to account for unobserved time-varying changes among advisors. Finally, $\kappa_{s-t}$ corresponds to state-year fixed effects to account for unobserved time-varying changes across states in which the bonds are issued. I cluster standard errors by state due to the segmented nature of the municipal bond market \citep{pirinsky2011market}. The results are robust to alternative dimensions of calculating standard errors including double clustering by state and year-month, as discussed in Section \ref{subsec:robustness} in the Internet Appendix.


\subsection{\label{subsec:selecting_nego}Identification Using Difference-in-Differences}

To justify using the difference-in-differences analysis for the causal implication in this setting, I compare treated (negotiated) and control (competitive) bonds. 
Figure \ref{fig:bond_chars} Panel A provides the kernel densities showing the distribution of primary market bond features like amount issued, coupon, offering yield and maturity between the treated and control groups (see description in Table \ref{table:vardescription}). Importantly, I find that the two groups look similar in the pattern of their distributions. As further validation, I show the quantile-quantile plots between treated and control bonds for these characteristics in Panel B of Figure \ref{fig:bond_chars} where most observations lie along the 45 degree slope. I also tabulate these characteristics between the treated and control bonds in Table \ref{table:summary_bonds} in the Internet Appendix. These evidence suggest that the two groups of bonds are similar in their overall distribution, supporting the strategy in Equation \eqref{eq:baseline}.


\begin{center}
    [INSERT FIGURE \ref{fig:bond_chars} ABOUT HERE]
\end{center}

Another obvious concern with the identification strategy above relates to selection. Issuers that choose to raise municipal debt via negotiation may be different from those who do not. In order to test whether selection may likely bias the estimates of a difference-in-differences design, I estimate a linear probability regression describing the choice of selling bonds via negotiation (indicator $\mathbbm{1}\{Nego._{b,i,t}$) as,
\begin{equation}\label{eq:like_nego}
 \mathbbm{1}\{Nego._{b,i,t}=1\} = \kappa_{t} + \beta*X_{b,i,t} + \epsilon_{b,i,t}       
\end{equation}
where subscript $b$ indicates the bond issuance, $i$ indicates the issuer, and $t$ indicates the time. $X_{b,i,t}$ includes variables controlling for issue size, average bond size, coupon, years to maturity, callable status, credit enhancement, insurance status, bank qualification, number of bonds, type of security (general obligation vs revenue), type of issuer, frequency of issuer borrowing, and fixed effects for rating group as well as use of funds (bond purpose). 

Using three approaches, I demonstrate how various factors affect the choice of negotiation. This analysis captures three years before the SEC Municipal Advisor Rule to focus on the ex-ante snapshot. Figure \ref{fig:like_nego} shows the estimated coefficients for each specification. When evaluating the observable characteristics \emph{Within Issuers}, I show that  none of them significantly determine the likelihood of negotiation. 


\begin{center}
    [INSERT FIGURE \ref{fig:like_nego} ABOUT HERE]
\end{center}


Moreover, I supplement this cross-sectional evidence with Figure \ref{fig:amt_nego} in the Internet Appendix showing the portion bonds sold via negotiation and competitive bidding. Nearly 40\% of the bond volume is consistently sold via negotiation around 2014. I also test this more formally in Figure \ref{fig:ll_nego} in the Internet Appendix by plotting the coefficients from regressing the likelihood of negotiation on half-year dummies after including issuer fixed effects. Taken together, these evidence mitigate concerns about selection into negotiation around the timing of the regulation.

\section{\label{sec:data}Data}
This paper uses municipal bonds data are from FTSE Russell (formerly known as Mergent) Municipal Database and the Municipal Securities Rulemaking Board (MSRB). Additionally, I hand-collect data on municipal advisor fees through Freedom Of Information Act (FOIA) requests made to state and local government bond issuers.
\noindent\subsection{\label{subsec:data_muniBonds}Municipal Bonds}
Municipal bond characteristics are obtained from the Municipal Bonds dataset by FTSE Russell (formerly known as Mergent MBSD). I retrieve the key bond characteristics such as CUSIP, dated date, the amount issued, size of the issue, offering type (method of sale), state of the issuing authority, name of the issuer, yield to maturity, tax status, insurance status, pre-refunding status, coupon rate, and maturity date for bonds. The baseline sample consists of fixed rate, tax-exempt bonds issued during January 2010 to December 2021. Issuers raised over USD 400 billion of municipal debt each year, mostly with advisors (Figure \ref{fig:total_iss}). I also use the average credit ratings for these bonds within one year of issuance. These CUSIP-level ratings are provided by S\&P, Moody's and Fitch. I encode character ratings into numerically equivalent values ranging from 28 for the highest quality to 1 for the lowest quality \citep{adelino2017economic}. 

\begin{center}
    [INSERT FIGURE \ref{fig:issuance_regis} ABOUT HERE]
\end{center}


The FTSE Russell database also provides the names of municipal advisor and underwriters involved in the bond issuance. Most bond issuances have a single municipal advisor. For a few cases with two advisors, I assign the issuance to the advisor with a larger cumulative volume advised. I use the lead manager as the underwriter. I manually check the names of advisors and underwriters for spelling errors and related data incongruities to standardize them during the sample period (see details in Section \ref{sec:adv_names}). \cite{bergstresser2018evolving} describe the evolution of the municipal advisor firm market. Figure \ref{fig:muni_regis} shows an increase in the number of municipal advising firms during 2010 to 2013, followed by a decline. The number of withdrawals by advisor firms peaks in 2014.Panel A of Table \ref{table:summary_bonds1} summarizes the top fifteen advisors in the sample and their relative share of volume advised.

\begin{center}
    [INSERT TABLE \ref{table:summary_bonds1} ABOUT HERE]
\end{center}

The FTSE Municipal Bonds dataset does not have the county name of each bond. I supplement this information from other sources like Bloomberg as in \cite{chava2024impact}. I use the Federal
Information Processing Standards (FIPS) code thus obtained to assign county level characteristics to bonds issued by local governments/issuers\endnote{Additionally, I define ``issuers'' based on the ultimate borrower identity from Bloomberg following \cite{gao2021good}. I obtain information on the type of (issuer) government i.e., state, city, county or other, from the Electronic Municipal Market Access (EMMA) data provided by the MSRB. I use the secondary market transactions data also from the MSRB database.}. The final sample comprises an average bond worth USD 2.7 million, issued at a coupon rate of 3.5\% with a maturity of ten years. Panel B of Table \ref{table:summary_bonds1} provides more details on the distribution of bond characteristics.

\noindent\subsection{\label{subsec:data_muniAdvFees}Municipal Advising and Underwriting Fees}
The municipal bond market has only recently been researched by academicians. It is not surprising that I do not find any commercial database that maintains a record of advisory fees charged by municipal advisors during issuance. To overcome this hurdle, I hand-collect this information by requesting these data under the FOIA at the sate-level\endnote{States vary substantially in their handling and maintenance of these records. Some states like CA, TX and WA had detailed information on the break up of fees paid to various (financial and legal) agents \citep{marlowe2013municipal} in each bond issuance within the state. This would include bonds issued by local governments as well as state agencies and authorities. In comparison, the state of New York was able to furnish information on the aggregate cost of issuance without providing a break-up to identify the fees paid to their municipal advisors. But New York City had more detailed information. Few other states (like IL and PA) only had information on their state level general obligation bonds. They denied collecting similar information from the local governments within their jurisdiction and guided that the request be made to each issuer separately.}. The United States municipal bond market has more than 50,000 unique bond issuers because even at the county level, different agencies may be issuing bonds separately. This makes the pursuit of gathering information by requesting each local issuer painstakingly time-consuming and infeasible. 


\begin{center}
    [INSERT FIGURE \ref{fig:muni_adv_fees2} ABOUT HERE]
\end{center}

Overall, I was able to gather data from 11 states corresponding to nearly USD 100 billion of municipal issuance each year during 2010-2021. This represents nearly one-fourth of new municipal bond issuance volume in each year during this period. Specifically, I obtain these data from: CA, TX, WA, FL, MD, PA, NM, RI, VT, LA, NY. Figure \ref{fig:muni_adv_fees2} shows the trend in municipal advisor fees for every USD 100 of municipal debt raised. Interestingly, I find little change in average advising fees during the period. This may be partially attributed to regulatory oversight. For example, SEC Commissioner Kara Stein noted in September 2013 that issuers may have faced excessive fees by municipal advisors\endnote{\href{https://www.sec.gov/news/statement/2013-09-18-open-meeting-statement-kms}{https://www.sec.gov/news/statement/2013-09-18-open-meeting-statement-kms}}. The SEC sought to address such problems by regulating the market for municipal advisors. Meanwhile, there is a decrease in underwriting fees (further discussed in Section \ref{subsec:mechanism} using underwriting spreads). Due to a lack of matching identifiers, it is not possible to link the advisor and underwriting fees data to the FTSE Municipal Bonds dataset described in Section \ref{subsec:data_muniBonds}.

\section{\label{sec:results}Results}
I discuss the baseline results in Section \ref{subsec:yield-results} for Equation \eqref{eq:baseline}, including raw evidence on municipal bond yield spreads and parallel pre-trends assumption. Section \ref{subsec:mechanism} focuses on the mechanism through underwriter spreads and the role of advisors in selecting underwriters. In Section \ref{subsec:hetero_size}, I discuss the heterogeneity by the size and sophistication of issuers. Finally, I examine the exit of municipal advisors in Section \ref{subsec:exitMA}. 


\subsection{\label{subsec:yield-results}Impact on Offering Yield Spreads of Local Governments}
In this Section, I begin by providing a graphical description of the raw offering yields (Section \ref{subsubsec:raw_yields}). Following this visual summary, I provide evidence from a dynamic difference-in-differences regression estimation in Section \ref{subsubsec:dyn_baseline}. Finally, I discuss the baseline result with the full set of controls and fixed effects in Section \ref{subsubsec:res_baseline}.

\subsubsection{\label{subsubsec:raw_yields}Raw Relationship in Offering Yields}
I start the analysis by a simple way of statistically summarizing the observed data: plotting the offering yields and the corresponding fitted curve, before and after the SEC Municipal Advisor Rule. Figure \ref{fig:bin_spreads} shows the binscatter of negotiated/``treated'' (circle) and competitive/``control'' (diamond) offering yields. I demarcated the promulgation of the Rule with a dashed vertical line. As shown, the yields tend to follow a downward trajectory until 2014 before the Rule. Thereafter, there is a slight increase in offering yields leading up to the SEC Rule. 

\begin{center}
    [INSERT FIGURE \ref{fig:bin_spreads} ABOUT HERE]
\end{center}

Importantly, I observe nearly parallel trends in the fitted curves for the treated and control bonds before the Rule. After the Rule in 2014, I observe a downward trend for treated yields resulting in a convergence of offering yields. Initially, there is some gap between the treated and control bonds right after the Rule (marked by the dashed vertical line). Negotiated yields tend to be higher than competitive yields; this is consistent with the literature \citep{robbins2007competition,guzman2012bonds,robbins2015missouri,liu2018effect,cestau2021should,bergstresser2023risk,gerrish2024meta}. However, the difference in offering yields between treated and control bonds reduces to nearly zero basis points by the middle of 2021. For clarity, I plot this difference in the shaded area at the base of the plot. Next, I follow this preliminary evidence with a robust analysis of the standard errors and regression outcomes for yields.



\subsubsection{\label{subsubsec:dyn_baseline}Dynamics in Difference-in-Differences Design}
Using the average offering yields in the primary market between the treated and control bonds during 2010 and 2021, I show the regression coefficients from the equation below:
\begin{equation} \label{eq:dyn_yield_CompNego}
y_{b,i,a,t} = \alpha + \delta_h * \sum_{h=2011H1}^{h=2019H1}Treated_{b,i,a} * Post_{h} + \beta_h * \sum_{h=2011H1}^{h=2019H1}Control_{b,i,a} * Post_{h}              +\eta_{i} + \epsilon_{b,i,a,t}
\end{equation}
where, $\eta_{i}$ represents issuer fixed effect and each coefficient $\delta_{h}$ corresponds to the twelve month periods ending June of that year. I estimate these time dummies for the treated and control bonds simultaneously, benchmarked to one year before the event window shown in Figure \ref{fig:did_spread}. Representing the yields from bond issuances on a twelve month scale ending June affords two advantages. First, I am clearly able to distinguish between the period before and after the SEC Municipal Advisor Rule, which became effective on July 1, 2014, and is depicted by the bold vertical line in the figure. Second, this frequency of representation is also consistent with the annual fiscal cycle of most local governments \citep{cuny2022information}.

\begin{center}
    [INSERT FIGURE \ref{fig:did_spread} ABOUT HERE]
\end{center}

The coefficients in panel (a) of Figure \ref{fig:did_spread} reveal a downward slope for offering yields, in general. This is not surprising given the monetary policy environment leading to lower yields in the financial markets. Importantly, before the SEC Rule, I find that the treated and control groups tend to follow nearly parallel trends with issuer fixed effects. This lends useful support to the main identification assumption \citep{bertrand2004much} that the treated group would follow the control group in the absence of the regulation. From July 2014 onward, I find that the offering yields for the treated bonds tend to decrease in comparison to the control bonds. Simultaneously, I plot the differences in coefficients over time along with their confidence intervals in panel (b) of Figure \ref{fig:did_spread}. The difference coefficients in the periods before the Rule (depicted by the bold vertical line) are nearly zero and statistically indistinguishable. After the SEC Rule, the yield coefficients are economically different from zero and statistically significant. 


The figure also reveals that the magnitude visibly increases after June 2016. Anecdotally, the timing of the higher impact on the dynamic coefficient of difference-in-difference in offering yields is also supported by the \href{https://www.sec.gov/news/press-release/2016-54}{SEC's enforcement against Kansas-based Central States Capital Markets in 2016}\endnote{I am thankful to an anonymous referee for this helpful suggestion.}. Central States agreed to settle the SEC’s charges by paying \$289,827.80 in disgorgement and interest and an \$85,000 civil penalty. Two of the firm's employees agreed to settle the charges by further paying a civil penalty each and agreeing to a bar from the financial services industry for some period. This was also the \href{https://www.proskauer.com/blog/first-sec-case-against-municipal-advisor-under-new-fiduciary-duty-03-16-2016}{first SEC case against a municipal advisor firm which imposed the new fiduciary duty obligation.}

\subsubsection{\label{subsubsec:res_baseline}Baseline Difference-in-Differences}
So far, I have visually summarized the raw relationship (via Figure \ref{fig:bin_spreads}) and demonstrated the parallel trends assumption (via Figure \ref{fig:did_spread}) in the difference-in-differences design. Now, I turn to the baseline effect on yields to quantify the magnitude due to the imposition of fiduciary duty by the Municipal Advisor Rule.

Table \ref{table:main_baseline} reports the main result using Equation \eqref{eq:baseline} to quantify the impact of advisors' fiduciary duty on yields. The coefficient of interest ($\beta_0$) represents the interaction term \textit{Treated $\times$ Post} corresponding to the difference in differences estimate. Extending the visual analysis, I first show the results in Column (1) with offering yields as the dependent variable. This model includes issuer fixed effect as well as state $\times$ year fixed effect to account for unobserved heterogeneity in estimating the coefficient within issuers, after controlling for time-varying factors at the state $\times$ year level. The effect is --10.58 basis points and is statistically significant. 

\begin{center}
    [INSERT TABLE \ref{table:main_baseline} ABOUT HERE]
\end{center}

Hereafter, I show results with offering yield spreads as the main dependent variable in Columns (2)-(5) by incrementally introducing additional controls. (See Table \ref{table:vardescription} for a description of key variables). First, Column (2) shows the same model as in Column (1) but changes the dependent variable to offering yield spreads. In Column (3), I include bond-level controls, rating fixed effects, and county-level controls. Specifically, I control for the coupon (\%); log(amount issued in \$); dummies for callable bonds, bond insurance, general obligation bond, bank qualification, refunding status, and credit enhancement; credit rating; remaining years to maturity; and inverse years to maturity. This model also controls for observable time-varying factors for the issuer based on the county-level economic conditions. Following \cite{gao2019financing}, I use the lagged values for log(labor force) and unemployment rate, and the percentage change in unemployment rate and labor force, respectively. 

It is also important to account for unobserved time-varying heterogeneity among advisors. In Columns (4)-(5), I further introduce advisor $\times$ year fixed effect. Column (4) shows that the difference-in-differences coefficient is --10.38 basis points and is statistically significant. The baseline specification corresponds to Column (5), where I include additional controls\endnote{I appreciate this valuable suggestion from an anonymous reviewer.} for the general interest rate environment (captured by changes in the Federal Funds rate and 10-year US Treasury yields) as well as the quality of underwriter (based on market share). Any effects due to advisor quality get absorbed in the advisor $\times$ year fixed effect. I find that the differential effect on treated bonds after the SEC Municipal Advisor Rule (represented by $\beta_0$) amounts to -10.62 basis points. In other words, negotiated bonds are issued at yields that are lower by 10.62 basis points\endnote{This magnitude is comparable to the cost of switching to negotiated sale (15--17 bps), when they are allowed \citep{cestau2021should}. This is also close to the effect of Affordable Care Act (ACA) on hospital bonds yield spreads (-13.6 bps) reported in \cite{gao2021good}. \cite{adelino2017economic} report an estimated reduction of 13--14 bps on offering yields due to rating re-calibration. Alternatively, \cite{kriz2017impact} report that municipal bond yield spreads change by 15 bps due to rating re-calibration. \cite{garrett2021conflicts} shows that the borrowing cost for local governments reduces by 11 bps due to MSRB Rule-G23 reform. Likewise, there is an impact of nearly 7 bps on some offering yields due to fiscal problems after municipal bankruptcy in \cite{yang2019negative}. \cite{cornaggia2022opioid} find that offering yields increase by 16.74 bps due to opioid abuse.} when compared to competitively issued bonds within the same issuer after the SEC Rule. 

This magnitude accounts for unobserved factors at the issuer level, as well as time-varying factors corresponding to the issuer's state. I also control for observable characteristics at the bond level and time-varying observed local economic conditions for the issuer based on their county. Further, I also absorb unobserved time-varying heterogeneity among advisors. For the average bond in the sample issued at a yield spread of 1.22\%, this means a reduction in yields of about 8.7\% (=10.62/122)\endnote{To understand the magnitude more closely, I offer a back-of-the-envelope calculation on interest cost savings due to reduced yield spreads \citep*{gao2021good}. On average, issuers raised negotiated bonds worth USD 70 million during the sample period. A reduction in yields of 10.62 basis points would amount to lower \textit{annual} interest cost of USD 74,000. This is nearly four times the per pupil expenditure by the average public elementary school in the United States during 2019-20 (see \href{https://nces.ed.gov/fastfacts/display.asp?id=66}{https://nces.ed.gov/fastfacts/display.asp?id=66}).}. In Table \ref{table:baseline_yld} in the Internet Appendix, I show similar results using offering yields as the dependent variable. Further, I consider several robustness checks to the baseline specification in Table \ref{table:robustness} of the Internet Appendix.


Overall, this section provides evidence supporting the identifying assumption for the parallel pre-trends in yields between the treated and control bonds. The baseline specification suggests that the yield spreads for negotiated bonds decreased by 10.62 basis points after the Rule. Next, I shed light on the mechanism.

\subsection{\label{subsec:mechanism}Mechanism}
In this Section, I first motivate the evidence from offering prices in Section \ref{subsubsec:offering_px} before showing the evidence on underwriter spreads and liquidity (Section \ref{subsubsec:underpricing_res}). Further, I explain my results in light of the role played by advisors in selecting underwriters in Section \ref{subsubsec:uw_intro}.

\subsubsection{\label{subsubsec:offering_px}Impact on Offering Price}
As further support for the main finding, I now provide evidence from offering prices as the dependent variable. Municipal bonds are usually priced on a face value of USD 100. In the absence of external monitoring, profit-maximizing underwriters may have incentives to price the municipal bonds below the market value. Specifically, underwriter profit increases in yields in the primary market issuance because it pays a lower price to the municipal issuer \citep{garrett2021conflicts}. They may be able to take advantage of the limited information possessed by issuers with respect to investor demand on specific municipal bonds. Such frictions from information asymmetry are likely to be higher for negotiated bonds over competitively bid bonds. 


Another motivation for underwriters to distort the offering price of municipal bonds could come from the prospect of future business. It is easier to sell low-priced securities to clients and they may reward the underwriter with future business \citep{liu2010economic}. Lowering the offering price may also enable underwriters to generate higher profits in selling bonds to investors subsequently \citep*{green2007dealer,green2007financial}. In this regard, I evaluate the observed offering price of municipal bonds at the time of issuance. Using the baseline specification in Equation \eqref{eq:baseline}, I report the results in Table \ref{table:baseline_px}.

\begin{center}
    [INSERT TABLE \ref{table:baseline_px} ABOUT HERE]
\end{center}

In Column (1), I show the results without any controls and include issuer and state $\times$ year fixed effects. The offering price on treated bonds increases by USD 1.85 (per USD 100 of face value of bond). As before, I introduce additional controls and fixed effects incrementally to take care of observed bond characteristics, county-level time-varying factors, interest rate environment, underwriter quality and unobserved factors across advisors. The final specification in Column (4) corresponds to the fully saturated difference-in-differences model showing an increase of USD 1.10 in the offering price for treated bonds after the Rule. The higher offering price implies greater dollar proceeds for issuers from the bond sale.


\subsubsection{\label{subsubsec:underpricing_res}Impact on Underwriter Spreads and Bond-level Liquidity}
In the bond market, researchers have evaluated factors affecting underwriting fees (\cite{luby2013empirical}) and whether the opacity of the market facilitates underpricing by financial intermediaries\endnote{Several papers have examined the underpricing of securities \citep*{welch1989seasoned,ljungqvist2003conflicts,ritter2003investment,eckbo2007security,green2007dealer,green2007financial} with a long literature in initial public offerings by underwriters in the equities market.}. This paper explains the lower offering yields by examining the municipal bond underwriter spreads after the imposition of fiduciary duty on municipal advisors. 


In a competitively sold bond, underwriters have to commit to the price submitted in the auction. But for negotiated sale, they arrive at the price based on discussions with the issuer. The pricing is not pre-determined and depends on demand from investors. With municipal advisors owing a fiduciary responsibility to issuers after the SEC regulation, issuers may be able to negotiate better with underwriters. Following \cite{cestau2013tax}, I compute the underwriter spreads by comparing the average price paid by customers ($P$) to the interdealer price ($V$), scaled by the interdealer price ($V$) and shown in basis points\endnote{I use bond transactions from the MSRB within the first month of trading for each bond.}. 




\begin{center}
    [INSERT TABLE \ref{table:underpricing} ABOUT HERE]
\end{center}


I follow the baseline specification in Equation \eqref{eq:baseline} and present the results in Table \ref{table:underpricing}. Specifically, Column (4) corresponds to the baseline specification with the full set of fixed effects and controls. I find that the underwriting spread reduces by nearly 13 basis points for the treated bonds after the SEC Rule. Broadly, the impact remains fairly stable across the various specifications and each of these estimates is statistically significant at the conventional levels of significance. Reduced underwriting spreads are also consistent with lower underwriting fees reported in the descriptive evidence in Figure \ref{fig:muni_adv_fees2} of Section \ref{subsec:data_muniAdvFees}. 


Taken together, the higher offering price and lower underwriting spreads explain the mechanism through which the fiduciary duty on municipal advisors benefits issuers. Under the new SEC Rule, municipal advisors have a greater incentive to align their interests with municipal issuers. Therefore, their better service to issuers results in negotiating higher offering price for the bonds. Such reduction in underpricing also reduces underwriting spreads.


Further, markups on investor purchases increase with the amount of interdealer trading before the trade \citep{schultz2012market} in the municipal bond market. Therefore, lower underpricing (lower underwriter spreads) among negotiated bonds may suggest that the bonds pass through fewer dealers before being held by investors. To test this hypothesis, I construct a measure of liquidity from post-issuance trades which is based on \cite{schwert2017municipal}\endnote{This measure is derived from \cite{jankowitsch2011price} and captures the dispersion of traded prices around the market ``consensus valuation''. Bond-level estimates of the price dispersion are obtained by taking the average of daily estimates within the first month of bond trading.}.

\begin{center}
    [INSERT TABLE \ref{table:baseline_disp} ABOUT HERE]
\end{center}

Table \ref{table:baseline_disp} shows the results using the main specification in Equation \eqref{eq:baseline}. As expected, I find that the average price dispersion decreases for negotiated bonds in comparison to competitively bid bonds after the Municipal Advisor Rule. I show the results by incrementally introducing controls and fixed effects from Column (1) through (4). The coefficient estimates are stable across each of these specifications and remain statistically significant throughout. Using the baseline specification in Column (4), these results suggest that the price dispersion reduces by about USD 0.03 for treated bonds after the regulation. This reduced liquidity measure is consistent with the mechanism explained previously.

\subsubsection{\label{subsubsec:uw_intro}Impact Due to Advisor's Ex-ante Role in ``Selecting'' Underwriter}
Advisors may also be entrusted with choosing an underwriting firm \citep{daniels2018does} based on their information. This becomes especially relevant for a negotiated sale because competitive bidding eliminates the requirement for pre-selection of the underwriting firm before the bond issuance. The principal-agent problem may arise from less-informed issuers engaging more informed advisors. In light of this, the decision to pre-select the underwriting firm in negotiated bonds becomes crucial. 


My analysis uses the ex-ante heterogeneity among issuers with respect to the role of advisors in choosing underwriters before the Rule. Ideally, I would want to use the precise information on advisors “selecting” underwriters. Without such data, I construct this measure by identifying whether the advisor introduces a new underwriter to the issuer for the first time. For example, when Alvord Union School District (CA) engaged Dolinka Grup Inc. as the municipal advisor in 2009, they ``\emph{introduced}'' Piper Jaffray \& Co. as an underwriter for the first time to this issuer. Previously, they had not worked with this underwriting firm\endnote{Other examples of this measure are provided in Table \ref{table:eg_byUWintro} in the Internet Appendix.}. 

\begin{center}
    [INSERT TABLE \ref{table:cs_byUWintro} ABOUT HERE]
\end{center}

In Table \ref{table:cs_byUWintro}, I analyze the ex-ante heterogeneity among issuers based on the average and weighted average likelihood of their advisor ``selecting'' an underwriter. This measure is high for issuers among whom advisors likely played a greater role in introducing new underwriters, as described above. Columns (3) and (6) correspond to the baseline specification with dummies interacted for the ex-ante classification of issuers. Additionally, this analysis also controls for group $\times$ year fixed effects. I find that the impact of fiduciary duty on lower offering yield spreads is driven by issuers where advisors play a greater (above median) role in selecting underwriters. In Column (3), the coefficient is --13.55 bps for the above median group and is statistically significant. Moreover, the difference in coefficients for the two groups is also significant. I find similar results using the weighted average measure in Column (6). The choice of underwriters is crucial with regard to the pricing of the bonds. Therefore, this evidence suggests that the imposition of fiduciary duty on advisors may drive greater yield reduction when advisors play a crucial role. 

Overall, this section explains the main finding of the paper. The reduction in offering yields is due to higher offering price and lower underwriting spreads. Finally, I show that the main result is driven by issuers among whom advisors likely play a greater role in ``selecting'' underwriters. The next section explores the heterogeneity in the baseline result.


\subsection{\label{subsec:hetero_size} Heterogeneity: Based on Size and Sophistication of Issuers}
So far, the evidence suggests that yield spreads reduce for negotiated bonds when compared to competitively bid bonds, after the Municipal Advisor Rule. Hand-collected data from FOIA requests indicates that the advisory fees paid by issuers on average have not gone up. Given these favorable factors, I first examine whether issuers are more likely to engage advisors. It is reasonable to expect that the prospect of lower yields \textit{without} incurring higher fees might encourage issuers to engage advisors more frequently \citep{vijayakumar2006role,allen2010does}.


Therefore, I evaluate the weighted average likelihood of recruiting an advisor \emph{within} issuers, on the extensive margin. Figure \ref{fig:ll_advised} shows the results from a linear regression of the average likelihood over annual (12-month) dummies with issuer fixed effect. I find that there is a 5\% increase in the likelihood of engaging an advisor after the SEC Municipal Advisor Rule. The omitted benchmark period corresponds to the half year at the start of the event window in 2010. 

\begin{center}
    [INSERT FIGURE \ref{fig:figure7} ABOUT HERE]
\end{center}

\cite{ang2017advance} find that issuers may accept NPV losses for short term cash flow savings. This may be especially important for smaller issuers. To understand more, I examine the likelihood of
engaging advisors based on the size of issuers. I present these results in Figure \ref{fig:ll_advised_bySize} by grouping issuers into small versus large, based on the size of their ex-ante municipal issuances. The evidence shows that the overall effect among all issuers is likely driven by large issuers only. In comparison, small issuers exhibit almost no change in their likelihood to engage advisors. 

The choice of engaging an advisor for a municipal bond issuance is endogenous to issuers. Issuers who are more likely to benefit from advised bonds may also be more likely to engage an advisor on the extensive margin. Revisiting the baseline analysis on advised bonds in Table \ref{table:cs_bySize}, I find that large (above median) issuers experience a greater reduction in yields for their treated bonds. The results use two approaches to sub-divide issuers based on their ex-ante average issuance (Columns (1)-(3)) and ex-ante median issuance (Columns (4)-(6)), respectively. Specifically, Columns (3) and (6) correspond to the baseline specification with the full set of controls and fixed effects. Additionally, I control for group $\times$ year fixed effect to account for unobserved time-varying heterogeneity among small versus large issuers. The impact on yield spreads is greater for large issuers by 12.60 bps (Column (3)) and 13.58 bps (Column (6)), respectively. Therefore, large issuers benefit more from the reduction in yield spreads.

\begin{center}
    [INSERT TABLE \ref{table:cs_bySize} ABOUT HERE]
\end{center}

As a direct consequence from differential impact on yield reduction based on issuer size, I analyze changes in municipal bond issuance after the SEC Municipal Advisor Rule. \cite{adelino2017economic} show that municipalities' financial constraints may impact the issuance of bonds. Similarly, \cite{boyson2022public} show that less wealthy school districts have difficulty obtaining municipal bond market funding. I classify issuers with greater (above median) reliance on negotiated bonds before the regulation as the “treated” group. The control group comprises issuers with below median reliance. Figure \ref{fig:new_issuance_ov} shows the overall results. Compared to the one year before the Rule,
treated issuers raise more municipal debt than control issuers. This is consistent with the ex-post reduction in offering yields which may increase the debt capacity for issuers. However, the analysis based on issuer size in Figure \ref{fig:new_issuance_bySize} shows that the increased issuance is primarily driven by large issuers. To reiterate, they also experience greater yield reduction. Meanwhile, there is a decrease in the amount of bonds raised by small issuers. 

\begin{center}
    [INSERT FIGURE \ref{fig:figure8} ABOUT HERE]
\end{center}




Larger issuers may also be more sophisticated in their bond issuance. I begin this analysis by identifying issuers that issued more complex bonds, ex-ante. The measure for complexity of bonds follows \cite{harris2006secondary,brancaccio2021search} to aggregate over six bond features: callable bonds, sinking fund provision, special redemption/extraordinary call features, nonstandard interest payment frequency, nonstandard interest accrual basis, and credit enhancement. 


The results show that the differential impact on yield spreads of issuers with above median complexity is 12.92 bps higher than those below median. This difference is statistically significant and economically meaningful. As before, it accounts for the average effect among low versus high complexity issuers by including group $\times$ year fixed effect. Similarly, Column (2) shows the results by focusing on weighted average complexity of advised bonds only to classify issuers. I find similar results as in Column (1), suggesting greater yield reduction among issuers with more complex bonds. 

\begin{center}
    [INSERT TABLE \ref{table:cs_bySophis} ABOUT HERE]
\end{center}

In Columns (3)-(5), I draw upon additional measures to quantify the ex-ante level of sophistication among issuers. First, in Column (3), I use the fraction of bonds with credit enhancement to represent the heterogeneity among issuers. Municipal issuers who are able to purchase more credit enhancement (usually include letters of credit and guarantees) are likely more sophisticated. Consistent with this, I find that yield spreads decrease by 14.75 bps for issuers with above median levels of credit enhancement.  Next, Column (4) uses the ex-ante average wage paid to the finance staff of local governments\endnote{This data is obtained from the US Census Bureau's Annual Survey of of Public Employment \& Payroll (ASPEP) for local governments.}. The results show that issuers with higher levels of wages for their finance staff benefit 7.51 bps more in terms of yield reduction. Finally, Column (5) shows similar results by using the ex-ante reliance on advised bonds. Issuers that are less reliant on advisors experience a greater reduction in yield spreads. 

Taken together, the evidence in this Section suggests that the baseline effect of reduced yields in negotiated bonds is driven by large and sophisticated issuers. As a result, larger issuers are also more likely to engage an advisor. The benefit from lower yield spreads increases debt issuance by large issuers, but small municipalities reduce their borrowing after the Rule.

\subsection{\label{subsec:exitMA}Advisor Fee Transparency and Advisor Exits}
The SEC imposed the fiduciary duty on municipal advisors through the Municipal Advisor Rule to address the principal-agent problem. The hand-collection of data on municipal advisor and underwriting fees across states in Section \ref{subsec:data_muniAdvFees} revealed US states that did or did not have data on such fees. This heterogeneity in state-level transparency on fee data through FOIA requests provides an interesting setting\endnote{I appreciate this indirect insight by an anonymous referee.}. States that did not respond with data for FOIA requests would likely pose a greater friction from the misalignment of incentives between the municipal advisor and issuer due to the principal-agent problem. This is premised on interpreting the lack of administrative data on fees paid to advisors and underwriters as a manifestation of the underlying friction. 

Table \ref{table:foia_st} shows the results from examining this heterogeneity among states. Columns (1) and (2) show sub-sample results by restricting the observations to states that did or did not have the FOIA data, respectively. Reduction in offering yield spreads is driven by states that did not have such data (Column (2)). This suggests that the Rule was instrumental in mitigating the principal-agent problem for issuers in states that did not have adequate governance controls to record the advisor fees data. In Column (3), I interact the baseline specification in Equation \eqref{eq:baseline} with dummies corresponding to the state in which the issuer belongs. The dummy corresponds to one where states provide data on advisor fees and these correspond to CA, TX, WA, FL, MD, PA, NM, RI, VT, LA, NY, and zero otherwise. This specification also includes the group $\times$ year fixed effect to account for unobserved time-varying heterogeneity within the two types of states. Column (3) shows that the baseline effect of reduction in yield spreads after the Rule is 13 bps more among states that did not have FOIA data. Stated differently, the analysis suggests that the benefit from the \emph{advice channel} is higher among states without FOIA data.

\begin{center}
    [INSERT TABLE \ref{table:foia_st} ABOUT HERE]
\end{center}

On the other hand, the \emph{fixed cost channel} from \citet{bhattacharya2019fiduciary} would suggest that some municipal advisors may exit the market after the Municipal Advisor Rule. This is likely due to the additional cost of compliance with the new regulatory requirements and increased paperwork \citep{bergstresser2018evolving}. I focus on the municipal advisors who advise on at least one issuance in each calendar year until June 2014. In Figure \ref{fig:numAdv_Regular}, I depict the number of regular advisors operating in the municipal bond market. The left axis shows that the number of regular advisors decreased by nearly 45\% from 206 in 2010 to 112 in 2021. Importantly, these municipal advisors worked on at least one issuance in each year before the regulation. Figure \ref{fig:numAdv_Regular} also shows the share of municipal bonds advised by these regular advisors on the right-hand axis. While these regular advisors worked on 90\% of the municipal bonds before the regulation, their share dropped to just over 75\% by 2021. 


\begin{center}
    [INSERT FIGURE \ref{fig:numAdv_Regular} ABOUT HERE]
\end{center}

In this context, I analyze the impact of advisors exiting the market on municipal bond yield spreads in Table \ref{table:exitingMA_over50}. In Column (1), I show results using Equation \eqref{eq:baseline} interacted with dummies corresponding to whether the issuer primarily depended on an exiting  advisor or not. I define issuers dependent on advisors when more than 50\% of their municipal debt issuance in the pre-period is advised by the exiting advisor\endnote{In untabulated results, I find that this is robust to using a lower threshold of 25\% of municipal debt issuance by the exiting advisor}. The results suggest that the reduction in offering yield spreads (--13.02 bps) is driven by issuers that do \emph{not} depend on an exiting advisor. 



\begin{center}
    [INSERT TABLE \ref{table:exitingMA_over50} ABOUT HERE]
\end{center}

However, the muted effect on issuers \emph{dependent} on exiting advisors masks the heterogeneity between small and large issuers. Columns (2) and (3) focus on the sub-samples of issuers that depend on an exiting advisor, ex-ante. Column (2) suggests that the yield spreads increase by 15.78 bps for small issuers, whereas the large issuers continue to experience a reduction in yield spreads (--12.09 bps) on their negotiated bonds after the MA Rule. The difference between the two groups is economically and statistically significant. Column (3) shows similar results even after including bonds issued without engaging municipal advisors for these issuers\endnote{It is possible that some issuers may raise debt without municipal advisors after the regulation.}. Overall, the evidence in this section shows that regulatory burden may result in the exit of some municipal advisors. This may increase the borrowing cost of small issuers that were dependent on these exiting advisors.


\section{\label{sec:conclusion}Conclusion}
I investigate how the imposition of fiduciary duty on municipal advisors affects municipal bond yields and advising fees. On the one hand, fiduciary duty may benefit municipal issuers by increasing the cost of poor advice by municipal advisors (\emph{advice channel}). However, the additional regulatory burden may also increase the cost of doing business (\emph{fixed cost channel}). It is unclear which of these two effects would dominate overall. I evaluate the overall implication of these two competing channels in the context of municipal bonds.

By focusing on the SEC Municipal Advisor Rule of 2014, I provide the first evidence on how fiduciary duty on municipal advisors affects municipal issuers. The findings suggest that the offering yield spreads on negotiated bonds reduced after the SEC Municipal Advisor Rule due to lower underwriting spreads. This is driven by issuers for whom advisors likely played a more significant ex-ante role in selecting underwriters. However, further analysis shows the heterogeneity between large and small issuers. 



\clearpage
\linespread{0.8}
	\newpage {\footnotesize
       \bibliography{Muni_adv_MFJ_R2}

\begin{thebibliography}{}

\bibitem[\protect\citeauthoryear{Adelino, Cunha, and Ferreira}{Adelino et~al.}{2017}]{adelino2017economic}
Adelino, M., I.~Cunha, and M.~A. Ferreira (2017).
\newblock The economic effects of public financing: Evidence from municipal bond ratings recalibration.
\newblock {\em The Review of Financial Studies\/}~{\em 30\/}(9), 3223--3268.

\bibitem[\protect\citeauthoryear{Allen and Dudney}{Allen and Dudney}{2010}]{allen2010does}
Allen, A. and D.~Dudney (2010).
\newblock Does the quality of financial advice affect prices?
\newblock {\em Financial Review\/}~{\em 45\/}(2), 387--414.

\bibitem[\protect\citeauthoryear{Anagol, Cole, and Sarkar}{Anagol et~al.}{2017}]{anagol2017understanding}
Anagol, S., S.~Cole, and S.~Sarkar (2017).
\newblock Understanding the advice of commissions-motivated agents: Evidence from the indian life insurance market.
\newblock {\em Review of Economics and Statistics\/}~{\em 99\/}(1), 1--15.

\bibitem[\protect\citeauthoryear{Ang, Green, Longstaff, and Xing}{Ang et~al.}{2017}]{ang2017advance}
Ang, A., R.~C. Green, F.~A. Longstaff, and Y.~Xing (2017).
\newblock Advance refundings of municipal bonds.
\newblock {\em The Journal of Finance\/}~{\em 72\/}(4), 1645--1682.

\bibitem[\protect\citeauthoryear{Azarmsa}{Azarmsa}{2021}]{azarmsa2021financing}
Azarmsa, E. (2021).
\newblock Financing infrastructure with inattentive investors: The case of us municipal governments.
\newblock {\em Available at SSRN 3945106\/}.

\bibitem[\protect\citeauthoryear{Babina, Jotikasthira, Lundblad, and Ramadorai}{Babina et~al.}{2021}]{babina2021heterogeneous}
Babina, T., C.~Jotikasthira, C.~Lundblad, and T.~Ramadorai (2021).
\newblock Heterogeneous taxes and limited risk sharing: Evidence from municipal bonds.
\newblock {\em The review of financial studies\/}~{\em 34\/}(1), 509--568.

\bibitem[\protect\citeauthoryear{Beatty, Gillette, Petacchi, and Weber}{Beatty et~al.}{2019}]{beatty2019rating}
Beatty, A., J.~Gillette, R.~Petacchi, and J.~Weber (2019).
\newblock Do rating agencies benefit from providing higher ratings? evidence from the consequences of municipal bond ratings recalibration.
\newblock {\em Journal of Accounting Research\/}~{\em 57\/}(2), 323--354.

\bibitem[\protect\citeauthoryear{Bergstresser and Herb}{Bergstresser and Herb}{2023}]{bergstresser2023risk}
Bergstresser, D. and P.~Herb (2023).
\newblock Do risk premia explain dealer markups in municipal bond offerings?
\newblock {\em Available at SSRN 3881297\/}.

\bibitem[\protect\citeauthoryear{Bergstresser and Luby}{Bergstresser and Luby}{2018}]{bergstresser2018evolving}
Bergstresser, D. and M.~J. Luby (2018).
\newblock The evolving municipal advisor market in the post dodd-frank era.
\newblock {\em Update\/}.

\bibitem[\protect\citeauthoryear{Bertrand, Duflo, and Mullainathan}{Bertrand et~al.}{2004}]{bertrand2004much}
Bertrand, M., E.~Duflo, and S.~Mullainathan (2004).
\newblock How much should we trust differences-in-differences estimates?
\newblock {\em The Quarterly journal of economics\/}~{\em 119\/}(1), 249--275.

\bibitem[\protect\citeauthoryear{Bhattacharya, Illanes, and Padi}{Bhattacharya et~al.}{2019}]{bhattacharya2019fiduciary}
Bhattacharya, V., G.~Illanes, and M.~Padi (2019).
\newblock Fiduciary duty and the market for financial advice.
\newblock Technical report, National Bureau of Economic Research.

\bibitem[\protect\citeauthoryear{Boyson}{Boyson}{2019}]{boyson2019worst}
Boyson, N.~M. (2019).
\newblock The worst of both worlds? dual-registered investment advisers.
\newblock {\em Dual-Registered Investment Advisers (December 1, 2019). Northeastern U. D’Amore-McKim School of Business Research Paper\/}~(3360537).

\bibitem[\protect\citeauthoryear{Boyson and Liu}{Boyson and Liu}{2022}]{boyson2022public}
Boyson, N.~M. and W.~Liu (2022).
\newblock Public bond issuance and education inequality.
\newblock {\em Available at SSRN 4099121\/}.

\bibitem[\protect\citeauthoryear{Brancaccio and Kang}{Brancaccio and Kang}{2021}]{brancaccio2021search}
Brancaccio, G. and K.~Kang (2021).
\newblock Search frictions and product design in the municipal bond market.

\bibitem[\protect\citeauthoryear{Bronshtein and Makridis}{Bronshtein and Makridis}{2020}]{bronshtein2020declining}
Bronshtein, G. and C.~A. Makridis (2020).
\newblock The declining insurance benefit in the municipal bond market.
\newblock {\em National Tax Journal\/}~{\em 73\/}(1), 115--156.

\bibitem[\protect\citeauthoryear{Bruno and Henisz}{Bruno and Henisz}{2024}]{bruno2024environmental}
Bruno, C.~C. and W.~J. Henisz (2024).
\newblock Environmental, social, and governance (esg) outcomes and municipal credit risk.
\newblock {\em Business \& Society\/}, 00076503231220541.

\bibitem[\protect\citeauthoryear{Cestau}{Cestau}{2020}]{cestau2020specialization}
Cestau, D. (2020).
\newblock Specialization investments and market power in the underwriting market for municipal bonds.

\bibitem[\protect\citeauthoryear{Cestau, Green, Hollifield, and Sch{\"u}rhoff}{Cestau et~al.}{2021}]{cestau2021should}
Cestau, D., R.~C. Green, B.~Hollifield, and N.~Sch{\"u}rhoff (2021).
\newblock Should state governments prohibit the negotiated sales of municipal bonds?
\newblock {\em Available at SSRN 3508342\/}.

\bibitem[\protect\citeauthoryear{Cestau, Green, and Sch{\"u}rhoff}{Cestau et~al.}{2013}]{cestau2013tax}
Cestau, D., R.~C. Green, and N.~Sch{\"u}rhoff (2013).
\newblock Tax-subsidized underpricing: The market for build america bonds.
\newblock {\em Journal of Monetary Economics\/}~{\em 60\/}(5), 593--608.

\bibitem[\protect\citeauthoryear{Chava, Malakar, and Singh}{Chava et~al.}{2022}]{cms2022_bankruptcy}
Chava, S., B.~Malakar, and M.~Singh (2022).
\newblock Communities as stakeholders: Impact of corporate bankruptcies on local governments.
\newblock {\em Working Paper\/}.

\bibitem[\protect\citeauthoryear{Chava, Malakar, and Singh}{Chava et~al.}{2024}]{chava2024impact}
Chava, S., B.~Malakar, and M.~Singh (2024).
\newblock Impact of corporate subsidies on borrowing costs of local governments: Evidence from municipal bonds.
\newblock {\em Review of Finance\/}~{\em 28\/}(1), 117--161.

\bibitem[\protect\citeauthoryear{Christoffersen, Evans, and Musto}{Christoffersen et~al.}{2013}]{christoffersen2013consumers}
Christoffersen, S.~E., R.~Evans, and D.~K. Musto (2013).
\newblock What do consumers’ fund flows maximize? evidence from their brokers’ incentives.
\newblock {\em The Journal of Finance\/}~{\em 68\/}(1), 201--235.

\bibitem[\protect\citeauthoryear{Cohen and Eappen}{Cohen and Eappen}{2015}]{cohen2015holds}
Cohen, N. and R.~Eappen (2015).
\newblock Who holds municipal bonds.
\newblock {\em Wells Fargo Municipal Commentary\/}.

\bibitem[\protect\citeauthoryear{Cohen, Mysak, Carney, and Zezas}{Cohen et~al.}{2017}]{cohen2017headline}
Cohen, N., J.~Mysak, S.~Carney, and M.~Zezas (2017).
\newblock Headline risk: Unexpected price changes and answering to the folks at the top.
\newblock {\em Municipal Finance Journal\/}~{\em 37\/}(4).

\bibitem[\protect\citeauthoryear{Cohen}{Cohen}{1989}]{cohen1989municipal}
Cohen, N.~R. (1989).
\newblock Municipal default patterns: An historical study.
\newblock {\em Public Budgeting \& Finance\/}~{\em 9\/}(4), 55--65.

\bibitem[\protect\citeauthoryear{Cohen}{Cohen}{2013}]{cohen2013municipal}
Cohen, N.~R. (2013).
\newblock Municipal bond insurance: Past, present, and future.
\newblock {\em Municipal Finance Journal\/}.

\bibitem[\protect\citeauthoryear{Cohen}{Cohen}{2017}]{cohen2017rating}
Cohen, N.~R. (2017).
\newblock Rating downgrades and retiree benefits.
\newblock {\em Municipal Finance Journal\/}~{\em 38\/}(1).

\bibitem[\protect\citeauthoryear{Cornaggia, Hund, Nguyen, and Ye}{Cornaggia et~al.}{2022}]{cornaggia2022opioid}
Cornaggia, K., J.~Hund, G.~Nguyen, and Z.~Ye (2022).
\newblock Opioid crisis effects on municipal finance.
\newblock {\em The Review of Financial Studies\/}~{\em 35\/}(4), 2019--2066.

\bibitem[\protect\citeauthoryear{Cuny, Li, Nakhmurina, and Watts}{Cuny et~al.}{2022}]{cuny2022information}
Cuny, C., K.~Li, A.~Nakhmurina, and E.~M. Watts (2022).
\newblock The information content of municipal financial statements: Large-sample evidence.
\newblock In {\em The Information Content of Municipal Financial Statements: Large-sample Evidence: Cuny, Christine| uLi, Ken| uNakhmurina, Anya| uWatts, Edward M.} [Sl]: SSRN.

\bibitem[\protect\citeauthoryear{Daniels, Dorminey, Smith, and Vijayakumar}{Daniels et~al.}{2018}]{daniels2018does}
Daniels, K., J.~Dorminey, B.~Smith, and J.~Vijayakumar (2018).
\newblock Does financial advisor quality improve liquidity and issuer benefits in segmented markets? evidence from the municipal bond market.
\newblock {\em Journal of Public Budgeting, Accounting \& Financial Management\/}.

\bibitem[\protect\citeauthoryear{Dimmock, Gerken, and Graham}{Dimmock et~al.}{2018}]{dimmock2018fraud}
Dimmock, S.~G., W.~C. Gerken, and N.~P. Graham (2018).
\newblock Is fraud contagious? coworker influence on misconduct by financial advisors.
\newblock {\em The Journal of Finance\/}~{\em 73\/}(3), 1417--1450.

\bibitem[\protect\citeauthoryear{Dimmock, Gerken, and Van~Alfen}{Dimmock et~al.}{2021}]{dimmock2021real}
Dimmock, S.~G., W.~C. Gerken, and T.~Van~Alfen (2021).
\newblock Real estate shocks and financial advisor misconduct.
\newblock {\em The Journal of Finance\/}~{\em 76\/}(6), 3309--3346.

\bibitem[\protect\citeauthoryear{Doty, Simpkins, Norwood, Singer, and Kommi}{Doty et~al.}{2018}]{doty2018regulation}
Doty, R., M.~Simpkins, L.~Norwood, N.~Singer, and L.~Kommi (2018).
\newblock Regulation and your consultants: Drawing the line.
\newblock {\em Municipal Finance Journal\/}~{\em 38\/}(4).

\bibitem[\protect\citeauthoryear{Eckbo, Masulis, and Norli}{Eckbo et~al.}{2007}]{eckbo2007security}
Eckbo, B.~E., R.~W. Masulis, and {\O}.~Norli (2007).
\newblock Security offerings.
\newblock {\em Handbook of empirical corporate finance\/}, 233--373.

\bibitem[\protect\citeauthoryear{Egan}{Egan}{2019}]{egan2019brokers}
Egan, M. (2019).
\newblock Brokers versus retail investors: Conflicting interests and dominated products.
\newblock {\em The Journal of Finance\/}~{\em 74\/}(3), 1217--1260.

\bibitem[\protect\citeauthoryear{Gaffney, Sanchez, Leifer, and Maas}{Gaffney et~al.}{2016}]{gaffney2016municipal}
Gaffney, S., D.~Sanchez, D.~Leifer, and T.~L. Maas (2016).
\newblock Municipal advisors: New standards and responsibilities.
\newblock {\em Municipal Finance Journal\/}~{\em 37\/}(3).

\bibitem[\protect\citeauthoryear{Gao, Lee, and Murphy}{Gao et~al.}{2019}]{gao2019financing}
Gao, P., C.~Lee, and D.~Murphy (2019).
\newblock Financing dies in darkness? {T}he impact of newspaper closures on public finance.
\newblock {\em Journal of Financial Economics\/}.

\bibitem[\protect\citeauthoryear{Gao, Lee, and Murphy}{Gao et~al.}{2021}]{gao2021good}
Gao, P., C.~Lee, and D.~Murphy (2021).
\newblock Good for your fiscal health? the effect of the affordable care act on healthcare borrowing costs.
\newblock {\em Journal of Financial Economics\/}.

\bibitem[\protect\citeauthoryear{Garrett}{Garrett}{2021}]{garrett2021conflicts}
Garrett, D. (2021).
\newblock Conflicts of interest in municipal bond advising and underwriting.
\newblock {\em Available at SSRN 3835504\/}.

\bibitem[\protect\citeauthoryear{Gerrish, Ivonchyk, Charles, Greer, and Moldogaziev}{Gerrish et~al.}{2024}]{gerrish2024meta}
Gerrish, E., M.~Ivonchyk, C.~Charles, R.~A. Greer, and T.~T. Moldogaziev (2024).
\newblock A meta-analysis of the state and local government borrowing costs.
\newblock {\em Public Administration Review\/}.

\bibitem[\protect\citeauthoryear{Green}{Green}{1993}]{green1993simple}
Green, R.~C. (1993).
\newblock A simple model of the taxable and tax-exempt yield curves.
\newblock {\em The Review of Financial Studies\/}~{\em 6\/}(2), 233--264.

\bibitem[\protect\citeauthoryear{Green, Hollifield, and Sch{\"u}rhoff}{Green et~al.}{2007a}]{green2007dealer}
Green, R.~C., B.~Hollifield, and N.~Sch{\"u}rhoff (2007a).
\newblock Dealer intermediation and price behavior in the aftermarket for new bond issues.
\newblock {\em Journal of Financial Economics\/}~{\em 86\/}(3), 643--682.

\bibitem[\protect\citeauthoryear{Green, Hollifield, and Sch{\"u}rhoff}{Green et~al.}{2007b}]{green2007financial}
Green, R.~C., B.~Hollifield, and N.~Sch{\"u}rhoff (2007b).
\newblock Financial intermediation and the costs of trading in an opaque market.
\newblock {\em The Review of Financial Studies\/}~{\em 20\/}(2), 275--314.

\bibitem[\protect\citeauthoryear{G{\"u}rkaynak, Sack, and Wright}{G{\"u}rkaynak et~al.}{2007}]{gurkaynak2007us}
G{\"u}rkaynak, R.~S., B.~Sack, and J.~H. Wright (2007).
\newblock The {US} treasury yield curve: 1961 to the present.
\newblock {\em Journal of Monetary Economics\/}~{\em 54\/}(8), 2291--2304.

\bibitem[\protect\citeauthoryear{Guzman and Moldogaziev}{Guzman and Moldogaziev}{2012}]{guzman2012bonds}
Guzman, T. and T.~Moldogaziev (2012).
\newblock Which bonds are more expensive? the cost differentials by debt issue purpose and the method of sale: An empirical analysis.
\newblock {\em Public Budgeting \& Finance\/}~{\em 32\/}(3), 79--101.

\bibitem[\protect\citeauthoryear{Harris and Piwowar}{Harris and Piwowar}{2006}]{harris2006secondary}
Harris, L.~E. and M.~S. Piwowar (2006).
\newblock Secondary trading costs in the municipal bond market.
\newblock {\em The Journal of Finance\/}~{\em 61\/}(3), 1361--1397.

\bibitem[\protect\citeauthoryear{Hazinski and Marlowe}{Hazinski and Marlowe}{2023}]{hazinski2022local}
Hazinski, T. and J.~Marlowe (2023).
\newblock Local lodging taxes during and after the pandemic.
\newblock {\em Municipal Finance Journal\/}.

\bibitem[\protect\citeauthoryear{Hildreth and Miller}{Hildreth and Miller}{2002}]{hildreth2002debt}
Hildreth, W.~B. and G.~J. Miller (2002).
\newblock Debt and the local economy: Problems in benchmarking local government debt affordability.
\newblock {\em Public Budgeting \& Finance\/}~{\em 22\/}(4), 99--113.

\bibitem[\protect\citeauthoryear{Hildreth and Zorn}{Hildreth and Zorn}{2005}]{hildreth2005evolution}
Hildreth, W.~B. and C.~K. Zorn (2005).
\newblock The evolution of the state and local government municipal debt market over the past quarter century.
\newblock {\em Public Budgeting \& Finance\/}~{\em 25\/}(4s), 127--153.

\bibitem[\protect\citeauthoryear{Inderst and Ottaviani}{Inderst and Ottaviani}{2012a}]{inderst2012competition}
Inderst, R. and M.~Ottaviani (2012a).
\newblock Competition through commissions and kickbacks.
\newblock {\em American Economic Review\/}~{\em 102\/}(2), 780--809.

\bibitem[\protect\citeauthoryear{Inderst and Ottaviani}{Inderst and Ottaviani}{2012b}]{inderst2012not}
Inderst, R. and M.~Ottaviani (2012b).
\newblock How (not) to pay for advice: A framework for consumer financial protection.
\newblock {\em Journal of Financial Economics\/}~{\em 105\/}(2), 393--411.

\bibitem[\protect\citeauthoryear{Jankowitsch, Nashikkar, and Subrahmanyam}{Jankowitsch et~al.}{2011}]{jankowitsch2011price}
Jankowitsch, R., A.~Nashikkar, and M.~G. Subrahmanyam (2011).
\newblock Price dispersion in otc markets: A new measure of liquidity.
\newblock {\em Journal of Banking \& Finance\/}~{\em 35\/}(2), 343--357.

\bibitem[\protect\citeauthoryear{Kriz and Xiao}{Kriz and Xiao}{2017}]{kriz2017impact}
Kriz, K.~A. and Y.~Xiao (2017).
\newblock The impact of rating recalibration on municipal bond yield spreads.
\newblock {\em Public Budgeting \& Finance\/}~{\em 37\/}(2), 83--101.

\bibitem[\protect\citeauthoryear{Landoni}{Landoni}{2018}]{landoni2018tax}
Landoni, M. (2018).
\newblock Tax distortions and bond issue pricing.
\newblock {\em Journal of Financial Economics\/}~{\em 129\/}(2), 382--393.

\bibitem[\protect\citeauthoryear{Liu}{Liu}{2018}]{liu2018effect}
Liu, G. (2018).
\newblock The effect of sale methods on the interest rate of municipal bonds: A heterogeneous endogenous treatment estimation.
\newblock {\em Public Budgeting \& Finance\/}~{\em 38\/}(2), 81--110.

\bibitem[\protect\citeauthoryear{Liu and Ritter}{Liu and Ritter}{2010}]{liu2010economic}
Liu, X. and J.~R. Ritter (2010).
\newblock The economic consequences of ipo spinning.
\newblock {\em The Review of Financial Studies\/}~{\em 23\/}(5), 2024--2059.

\bibitem[\protect\citeauthoryear{Ljungqvist}{Ljungqvist}{2003}]{ljungqvist2003conflicts}
Ljungqvist, A. (2003).
\newblock Conflicts of interest and efficient contracting in ipos.
\newblock {\em NYU, Ctr for Law and Business Research Paper\/}~(03-03).

\bibitem[\protect\citeauthoryear{Longstaff, Mithal, and Neis}{Longstaff et~al.}{2005}]{longstaff2005corporate}
Longstaff, F.~A., S.~Mithal, and E.~Neis (2005).
\newblock Corporate yield spreads: Default risk or liquidity? {N}ew evidence from the credit default swap market.
\newblock {\em The Journal of Finance\/}~{\em 60\/}(5), 2213--2253.

\bibitem[\protect\citeauthoryear{Luby and Moldogaziev}{Luby and Moldogaziev}{2013}]{luby2013empirical}
Luby, M. and T.~Moldogaziev (2013).
\newblock An empirical examination of the determinants of municipal bond underwriting fees.
\newblock {\em Municipal finance journal\/}~{\em 34\/}(2), 13--50.

\bibitem[\protect\citeauthoryear{Luby}{Luby}{2012}]{luby2012federal}
Luby, M.~J. (2012).
\newblock Federal intervention in the municipal bond market: The effectiveness of the build america bond program and its implications on federal and subnational budgeting.
\newblock {\em Public Budgeting \& Finance\/}~{\em 32\/}(4), 46--70.

\bibitem[\protect\citeauthoryear{Luby and Orr}{Luby and Orr}{2019}]{luby2019nic}
Luby, M.~J. and P.~Orr (2019).
\newblock From nic to tic to ray: Estimating lifetime cost of capital for municipal borrowers.
\newblock {\em Municipal Finance Journal\/}~{\em 39\/}(4).

\bibitem[\protect\citeauthoryear{Marlowe}{Marlowe}{2007}]{marlowe2007much}
Marlowe, J. (2007).
\newblock Much ado about nothing? the size and credit quality implications of municipal other postemployment benefit liabilities.
\newblock {\em Public Budgeting \& Finance\/}~{\em 27\/}(2), 104--131.

\bibitem[\protect\citeauthoryear{Marlowe}{Marlowe}{2009}]{marlowe2009method}
Marlowe, J. (2009).
\newblock Method of sale, price volatility, and borrowing costs on new issue municipal bonds.
\newblock In {\em Method of Sale, Price Volatility, and Borrowing Costs on New Issue Municipal Bonds: Marlowe, Justin}. [Sl]: SSRN.

\bibitem[\protect\citeauthoryear{Marlowe}{Marlowe}{2013}]{marlowe2013municipal}
Marlowe, J. (2013).
\newblock Municipal bond liquidity before and after the financial crisis.
\newblock {\em Available at SSRN 2206730\/}.

\bibitem[\protect\citeauthoryear{Marlowe}{Marlowe}{2020}]{marlowe2020municipal}
Marlowe, J. (2020).
\newblock Do municipal bond exchange-traded funds improve market quality?
\newblock {\em Hutchins Center Working Papers\/}.

\bibitem[\protect\citeauthoryear{Mughan}{Mughan}{2021}]{mughan2021municipal}
Mughan, S. (2021).
\newblock Municipal reliance on fine and fee revenues: How local courts contribute to extractive revenue practices in us cities.
\newblock {\em Public Budgeting \& Finance\/}~{\em 41\/}(2), 22--44.

\bibitem[\protect\citeauthoryear{Neumann}{Neumann}{2022}]{neumann2022municipal}
Neumann, J. (2022).
\newblock Municipal debt and the equity cross-section of states.
\newblock {\em Available at SSRN 4101404\/}.

\bibitem[\protect\citeauthoryear{Painter}{Painter}{2020}]{painter2020inconvenient}
Painter, M. (2020).
\newblock An inconvenient cost: The effects of climate change on municipal bonds.
\newblock {\em Journal of Financial Economics\/}~{\em 135\/}(2), 468--482.

\bibitem[\protect\citeauthoryear{Park, Matkin, and Marlowe}{Park et~al.}{2017}]{park2017internal}
Park, Y.~J., D.~S. Matkin, and J.~Marlowe (2017).
\newblock Internal control deficiencies and municipal borrowing costs.
\newblock {\em Public Budgeting \& Finance\/}~{\em 37\/}(1), 88--111.

\bibitem[\protect\citeauthoryear{Pirinsky and Wang}{Pirinsky and Wang}{2011}]{pirinsky2011market}
Pirinsky, C.~A. and Q.~Wang (2011).
\newblock Market segmentation and the cost of capital in a domestic market: Evidence from municipal bonds.
\newblock {\em Financial Management\/}~{\em 40\/}(2), 455--481.

\bibitem[\protect\citeauthoryear{Ritter}{Ritter}{2003}]{ritter2003investment}
Ritter, J.~R. (2003).
\newblock Investment banking and securities issuance.
\newblock In {\em Handbook of the Economics of Finance}, Volume~1, pp.\  255--306. Elsevier.

\bibitem[\protect\citeauthoryear{Rizzi}{Rizzi}{2022}]{rizzi2022nature}
Rizzi, C. (2022).
\newblock Nature as a defense from disasters: Natural capital and municipal bond yields.
\newblock {\em Available at SSRN 4038371\/}.

\bibitem[\protect\citeauthoryear{Robbins and Simonsen}{Robbins and Simonsen}{2007}]{robbins2007competition}
Robbins, M.~D. and B.~Simonsen (2007).
\newblock Competition and selection in municipal bond sales: Evidence from missouri.
\newblock {\em Public Budgeting \& Finance\/}~{\em 27\/}(2), 88--103.

\bibitem[\protect\citeauthoryear{Robbins and Simonsen}{Robbins and Simonsen}{2015}]{robbins2015missouri}
Robbins, M.~D. and B.~Simonsen (2015).
\newblock Missouri municipal bonds: The cost of no reforms.
\newblock {\em Municipal Finance Journal\/}~{\em 36\/}(1).

\bibitem[\protect\citeauthoryear{Schultz}{Schultz}{2012}]{schultz2012market}
Schultz, P. (2012).
\newblock The market for new issues of municipal bonds: The roles of transparency and limited access to retail investors.
\newblock {\em Journal of Financial Economics\/}~{\em 106\/}(3), 492--512.

\bibitem[\protect\citeauthoryear{Schwert}{Schwert}{2017}]{schwert2017municipal}
Schwert, M. (2017).
\newblock Municipal bond liquidity and default risk.
\newblock {\em The Journal of Finance\/}~{\em 72\/}(4), 1683--1722.

\bibitem[\protect\citeauthoryear{Smull, Kodra, Stern, Teras, Bonanno, and Doyle}{Smull et~al.}{2023}]{smull2023climate}
Smull, E., E.~Kodra, A.~Stern, A.~Teras, M.~Bonanno, and M.~Doyle (2023).
\newblock Climate, race, and the cost of capital in the municipal bond market.
\newblock {\em Plos one\/}~{\em 18\/}(8), e0288979.

\bibitem[\protect\citeauthoryear{Vijayakumar and Daniels}{Vijayakumar and Daniels}{2006}]{vijayakumar2006role}
Vijayakumar, J. and K.~N. Daniels (2006).
\newblock The role and impact of financial advisors in the market for municipal bonds.
\newblock {\em Journal of Financial Services Research\/}~{\em 30}, 43--68.

\bibitem[\protect\citeauthoryear{Welch}{Welch}{1989}]{welch1989seasoned}
Welch, I. (1989).
\newblock Seasoned offerings, imitation costs, and the underpricing of initial public offerings.
\newblock {\em The Journal of Finance\/}~{\em 44\/}(2), 421--449.

\bibitem[\protect\citeauthoryear{White}{White}{2014}]{white2014}
White, M. (2014).
\newblock Testimony on ``{O}versight of {F}inancial {S}tability and {D}ata {S}ecurity''.
\newblock {\em United States Senate Committee on Banking, Housing, and Urban Affairs\/}, Accessed August 2022 from: \href{https://www. sec.gov/News/Testimony/Detail/Testimony/1370540757488}{https://www. sec.gov/News/Testimony/Detail/Testimony/1370540757488}.

\bibitem[\protect\citeauthoryear{Yang}{Yang}{2017}]{yang2017financial}
Yang, L. (2017).
\newblock Financial management conservatism under constraints: Tax and expenditure limits and local deficit financing during the great recession.
\newblock {\em Local Government Studies\/}~{\em 43\/}(6), 946--965.

\bibitem[\protect\citeauthoryear{Yang}{Yang}{2019}]{yang2019negative}
Yang, L. (2019).
\newblock Negative externality of fiscal problems: Dissecting the contagion effect of municipal bankruptcy.
\newblock {\em Public Administration Review\/}~{\em 79\/}(2), 156--167.

\bibitem[\protect\citeauthoryear{Yang}{Yang}{2021}]{yang2021auditor}
Yang, L. (2021).
\newblock Auditor or adviser? auditor (in) dependence and its impact on financial management.
\newblock {\em Public Administration Review\/}~{\em 81\/}(3), 475--487.

\bibitem[\protect\citeauthoryear{Yang and Abbas}{Yang and Abbas}{2020}]{yang2020general}
Yang, L. and Y.~Abbas (2020).
\newblock General-purpose local government defaults: Type, trend, and impact.
\newblock {\em Public Budgeting \& Finance\/}~{\em 40\/}(4), 62--85.

\bibitem[\protect\citeauthoryear{Yang and Winecoff}{Yang and Winecoff}{2022}]{yang2022municipal}
Yang, L. and R.~Winecoff (2022).
\newblock Municipal bond sectoral risk and information intermediation in uncertain times: Evidence from the covid-19 pandemic.
\newblock {\em Public Budgeting \& Finance\/}~{\em 42\/}(4), 34--53.

\end{thebibliography}
		\bibliographystyle{chicago}
}

\vspace{2em}


\theendnotes


\clearpage
\newpage
\onehalfspacing
\setcounter{table}{0}
\renewcommand{\thetable}{A\arabic{table}}
\setcounter{figure}{0}
\renewcommand{\thefigure}{A\arabic{figure}}
\begin{center}
\textbf{Table A1:} Description of Key Variables\\
\end{center}
\vspace{-3em}
\begin{xltabular}{\linewidth}{ l X c}
\caption*{\\
\footnotesize{This table reports variable definitions. Data sources include the municipal bond transaction data from the Municipal Securities Rulemaking Board (MSRB), FTSE Russell's Municipal Bond Securities Database (FTSE, formerly known as Mergent MBSD), zero coupon yield provided by Finance and Economics Discussion Series (FEDS), and macroeconomic interest rate variables from the Federal Reserve Bank of St. Louis (FRB-SL).\\}
\vspace{-1em}
}
\label{table:vardescription}\\
\toprule
 Variable & Description  & Source\\
\midrule
\endfirsthead

\toprule
 Variable &  Description  & Source \\
\midrule
\endhead
\bottomrule
\endfoot
 
\emph{Treated} & Dummy set to one for bonds sold via negotiation. This dummy equals zero for competitively bid bonds. & FTSE
\\
\emph{Post} & Dummy that is assigned a value of one for time after SEC Municipal Advisor Rule became effective on July 1, 2014, and zero otherwise. & SEC
\\
\emph{GO Bond Dummy} & Dummy variable for general obligation bond. A GO bond is a municipal bond backed by the credit and taxing power of the issuing jurisdiction rather than the revenue from a given project. & FTSE
\\
\emph{Log(Amount)} & Log transformation of the dollar amount of the individual bond's (9-digit CUSIP) original offering. & FTSE
\\
\emph{Callable Dummy} & Dummy variable that equals 1 if the issue is callable and is 0 otherwise. & FTSE
\\
\emph{Insured Dummy} & Dummy variable that equals 1 if the issue is insured and is 0 otherwise. & FTSE
\\
\emph{Remaining Maturity} & Individual bond maturity measured in years. & FTSE
\\
\emph{Inverse Maturity} & Inverse of the value of \emph{Remaining Maturity}; to account for non-linearity. & FTSE
\\
\emph{Offering Price} & The price, expressed as a percentage of par, at which the bond was originally sold to investors & FTSE
\\
\emph{Offering Yield} & Yield to maturity at the time of issuance for a given bond (CUSIP), based on the coupon and any discount or premium to par value at the time of sale. & FTSE
\\
\emph{Yield Spread} & The difference between the \emph{Offering Yield} and the coupon-equivalent risk-free yield ($r_t$). The risk-free yield is based on the present value of coupon payments and the face value of the municipal bond using the US treasury yield curve based on maturity-matched zero-coupon yields by \cite*{gurkaynak2007us}. This yield spread calculation is similar to \cite*{longstaff2005corporate}. & FTSE, FEDS
\\
\emph{Underwriter Spread} & Trade-weighted difference between the average price paid by customers to buy the bonds ($P$) versus the interdealer price ($V$), scaled by the interdealer price ($V$) and shown in basis points \citep*{cestau2013tax}. & FTSE, MSRB
\\
\emph{Price Dispersion} & Average dispersion using traded prices around the market ``consensus valuation'' using \cite{schwert2017municipal}, based on \citet*{jankowitsch2011price}. Bond-level estimates of the price dispersion measure obtained by taking the average of daily estimates in the first month of trading. & FTSE, MSRB
\\
\emph{Rating} & Numeric value corresponding to the bond's credit rating from S\&P, Moody's or Fitch. We use ratings within one year of bond issuance. Following \cite{adelino2017economic}, we map the ratings into numeric values where the lowest rating is assigned the value of one and the higher ratings are assigned higher numeric values, progressively. & FTSE
\\
\emph{UW Mkt. Share(\%) } & It is the share of municipal bond volume underwritten by each underwriter in a year, divided by the total municipal debt issued in that year.  & FTSE
\\
\emph{$\triangle$Fed Funds$_{t-1}$} & Captures the rolling monthly lag of the quarterly change in Federal Funds rate. & FRB-SL
\\
\emph{$\triangle$UST10Y$_{t-1}$} & Captures the rolling monthly lag of the semi-annual change in US 10-year Treasury yields. & FRB-SL
\\
\end{xltabular}


\clearpage


\setcounter{table}{0}
\renewcommand{\thetable}{\arabic{table}}

\setcounter{figure}{0}
\renewcommand{\thefigure}{\arabic{figure}}

\clearpage

\begin{figure*}
\raggedright\textbf{Panel A: Distribution of Municipal Bond Characteristics (2010-2021)}\\
    \centering
\includegraphics[scale=1.25]{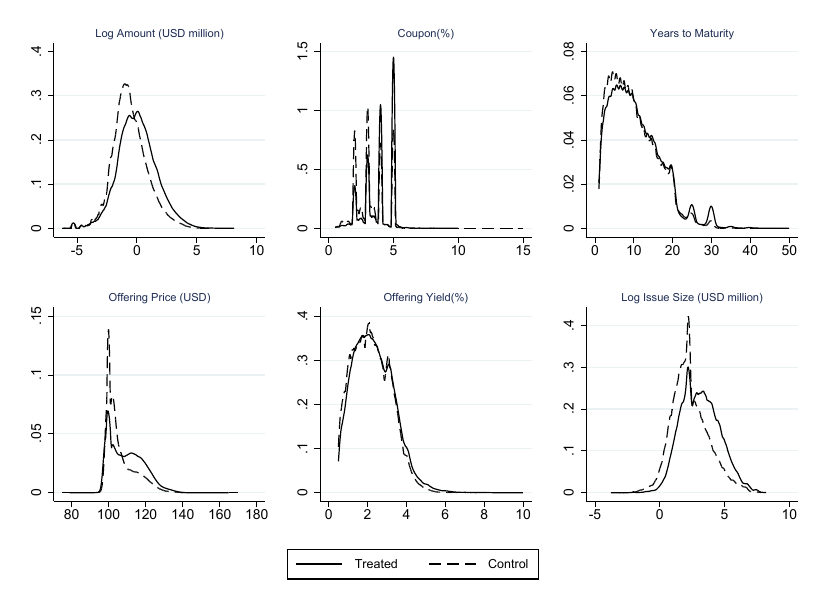}   
\label{fig:bond_prim_chars}
    \phantomcaption
\end{figure*}

\begin{figure*}
\raggedright\textbf{Panel B: Quantile-Quantile plot of Municipal Bond Characteristics (2010-2021)}\\
\ContinuedFloat
    \centering
\includegraphics[scale=0.75]{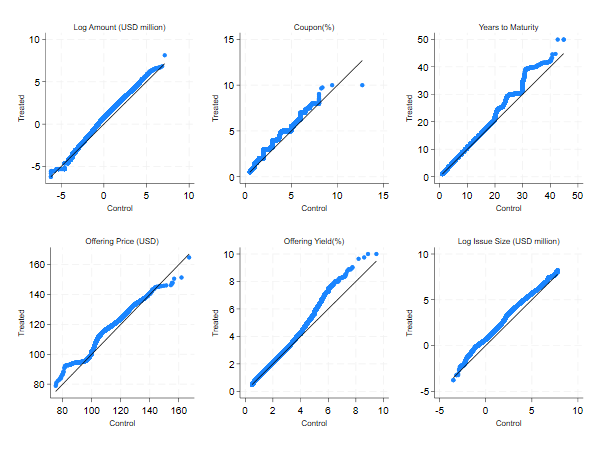}
\label{fig:qq_chars}
\caption{\textbf{Municipal Bond Characteristics}: \footnotesize{The figure shows the primary market characteristics of bonds issued with advisors at the time of issuance. Bonds sold via negotiated sale consist of the ``treated'' group, whereas competitively auctioned bonds comprise the ``control'' set. The sample focuses on fixed rate, tax-exempt bonds issued during 2010-2021. Panel B shows the quantile-quantile plot for these characteristics between treated and control bonds.}}
     \label{fig:bond_chars}
\end{figure*}

\begin{figure}
\begin{center}
\textbf{Linear Probability Estimates Explaining Choice of Negotiation}
\end{center}
\centering
\includegraphics[scale=1.2]{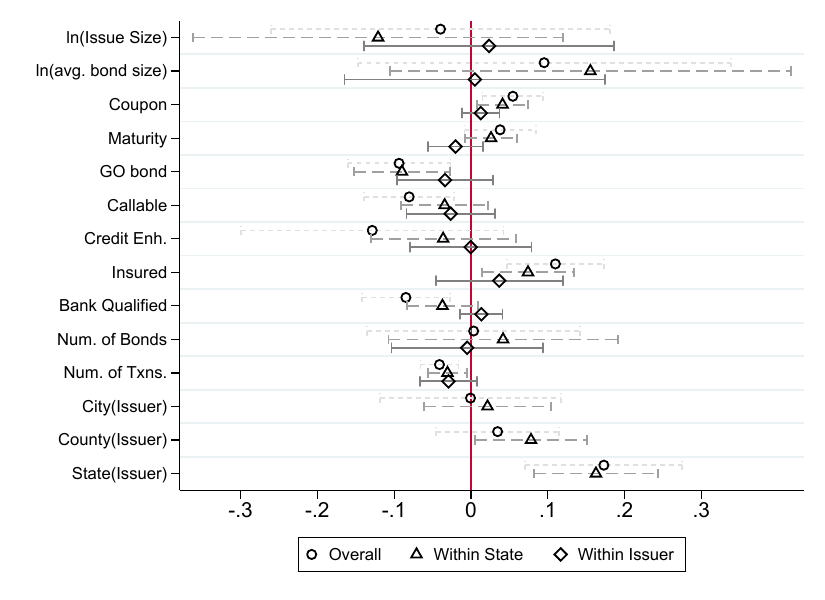}
\caption{\textbf{Linear Probability Estimates Explaining Choice of Negotiation}: \footnotesize{The figure shows the point estimates and 95\% confidence intervals using Equation \eqref{eq:like_nego} regressing the choice of negotiation on issuer and bond issue characteristics. The sample focuses on the three years before the SEC Municipal Advisor Rule to capture the ex-ante snapshot. Characteristics with continuous measurements are normalized to standard deviations. \emph{Overall} balance shows the estimates without including any geographic controls linked to the issuer. Next, \emph{Within State} balance corresponds to the estimates after including state fixed effects for the issuer. Finally, \emph{Within Issuer} shows results obtained from including issuer fixed effects.}}
\label{fig:like_nego}
\end{figure}

\begin{figure}
    \raggedright\textbf{Total Primary Market Issuance in Municipal Bonds}\\
    \begin{subfigure}{\linewidth}
    \centering
\includegraphics[scale=0.7]{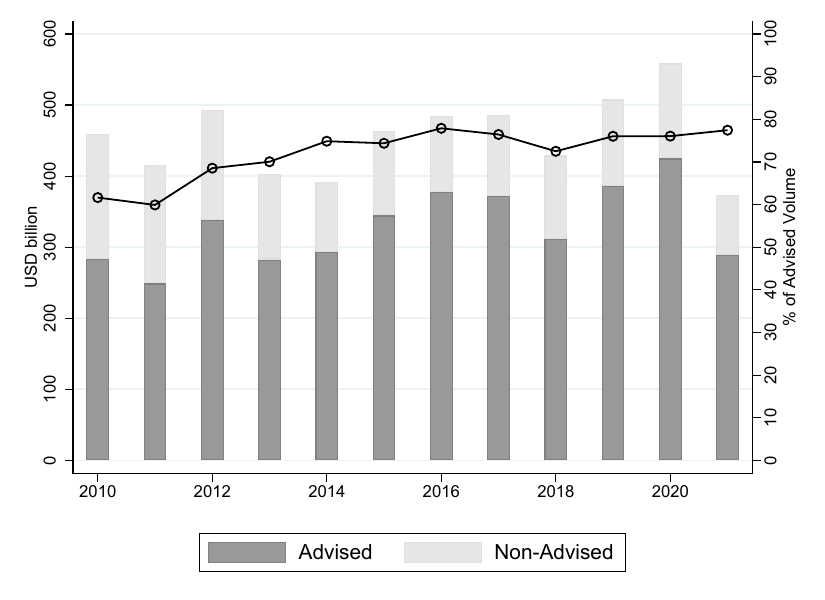}
    \caption{} 
    \label{fig:total_iss}
    \end{subfigure}
    
    \medskip
    \raggedright\textbf{Registration Activity of Municipal Advisor (MA) firms}\\
    \begin{subfigure}{\linewidth}
    \centering
\includegraphics[scale=0.7]{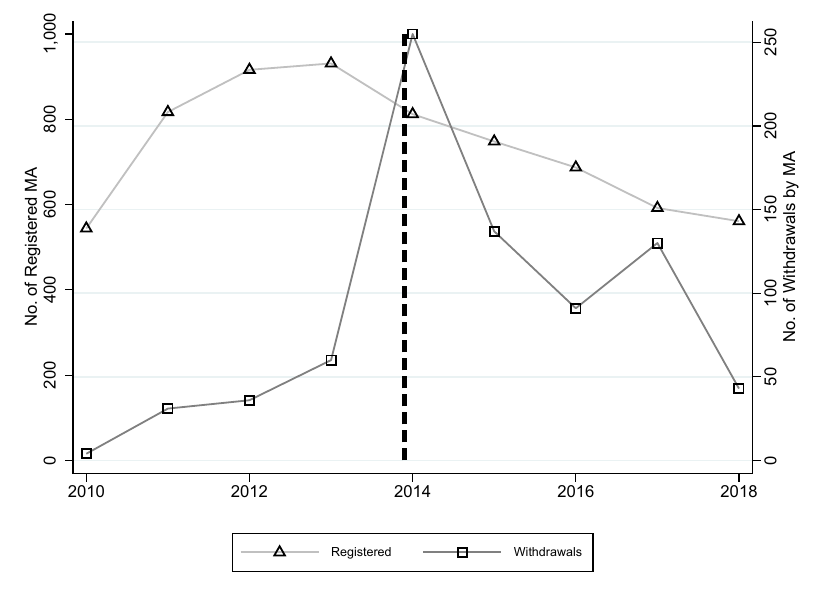}
    \caption{} 
    \label{fig:muni_regis}
    \end{subfigure}

\caption{\textbf{Municipal Market Issuance and Municipal Advisor Registration}: \footnotesize{In this figure, I show the issuance and registration activity in the municipal bond market. Panel (a) shows the total volume (in USD billion) of municipal debt issued during 2010-2021 from the FTSE Russell Municipal Bonds database in the primary market on the left axis. I also show the split based on advised versus non-advised bonds using the vertical bars. The line graph corresponds to the right axis, showing the percentage of advised bonds. Panel (b) shows the registration activity by municipal advisor (MA) firms during 2010-2018. This information is obtained from Table 1 in  \citet*{bergstresser2018evolving}. The left axis reports the number of firms registered as municipal advisors. The right axis provides the number of withdrawals filed by MA firms during this period.}}
\label{fig:issuance_regis}
\end{figure}

\begin{figure}
\begin{center}
\textbf{Municipal Advising and Underwriting Fees}
\end{center}
\centering
\includegraphics[scale=1.25]{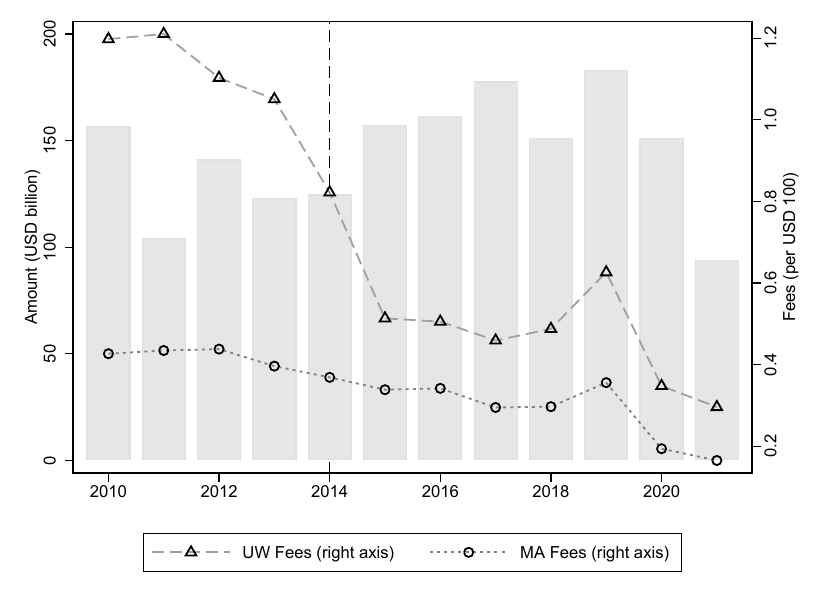}
\caption{\textbf{Municipal Advising and Underwriting Fees:} 
\footnotesize{The figure shows the municipal advising and underwriting fees paid by issuers alongside the corresponding amount of municipal debt raised. The vertical bars show the aggregated amount of municipal debt issued by state and local governments on the left axis for which I have the fees information during 2010-2021. The connected line with circles depicts the fees paid to municipal advisors for every USD 100 of municipal debt raised on the right axis. Likewise, the connected line with triangles depicts the fees paid to underwriters for every USD 100 of municipal debt raised on the right axis. These data were obtained under FOIA requests from 11 states: CA, TX, WA, FL, MD, PA, NM, RI, VT, LA, NY. See Section \ref{subsec:data_muniAdvFees} for details.}}
\label{fig:muni_adv_fees2}
\end{figure}

\begin{figure}
\begin{center}
\textbf{Raw Binscatter of Offering Yields (basis points)}
\end{center}
\centering
\includegraphics[scale=1.2]{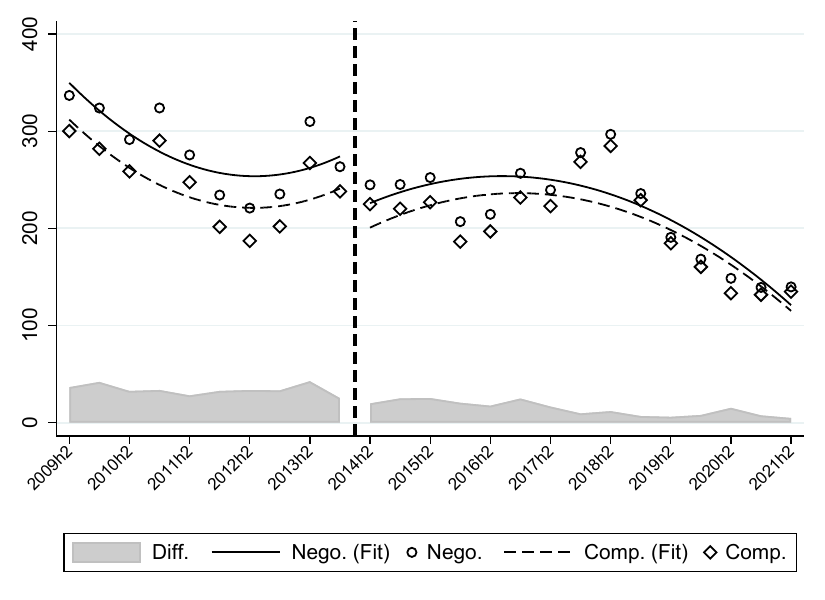}
\caption{\textbf{Binscatter of Offering Yields:} \footnotesize{This figure reports the binscatter of offering yield in basis points on the y-axis. Each dot is the average of the yields calculated on every $5^{th}$ percentile of the sample for treated and control bonds, separately. The figure also shows the corresponding fitted lines for the negotiated and competitively bid bonds. The difference between the two groups is represented in the shaded portion. The observed pattern suggests a parallel trend until June 2014, followed by a downward trend of spreads in the treated bonds after the SEC Municipal Advisor Rule. This analysis should be interpreted in a non-causal way, as no fixed effects and controls are included.}}
\label{fig:bin_spreads}
\end{figure}

\begin{figure}
    \raggedright\textbf{Offering Yields (in basis points)}\\
    \begin{subfigure}{\linewidth}
    \centering
    \includegraphics[scale=.85]{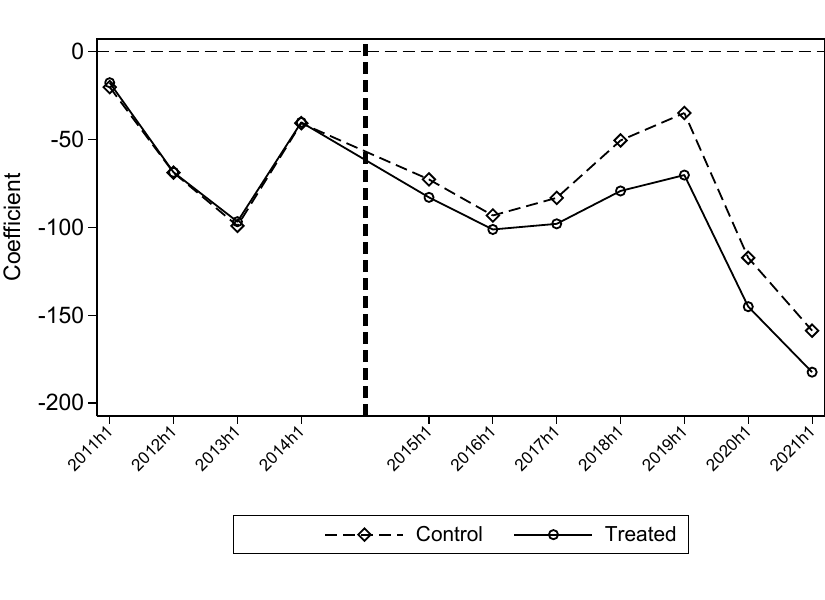}
    
    \caption{} 
    \end{subfigure}
    
    \medskip
    \raggedright\textbf{Difference in Yields (in basis points)}\\
    \begin{subfigure}{\linewidth}
    \centering
    \includegraphics[scale=.85]{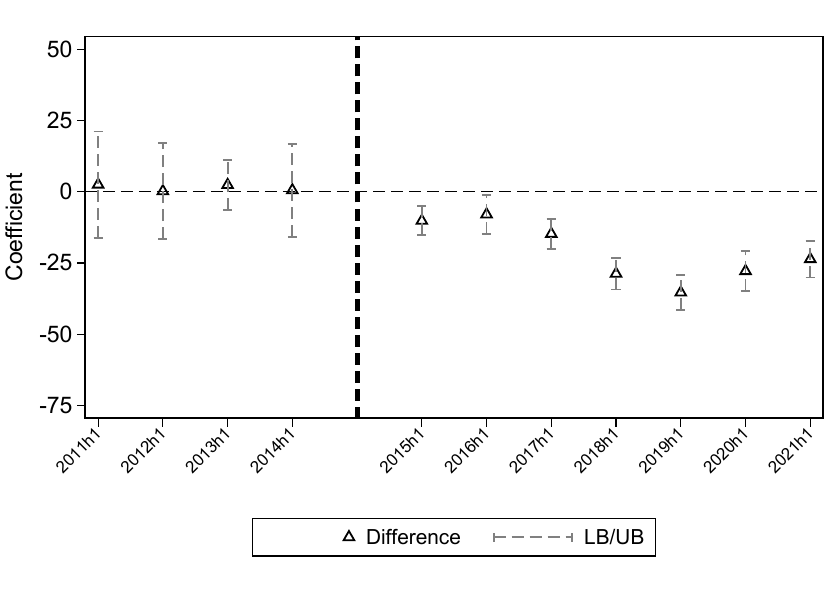}
    \caption{} 
    \end{subfigure}
    
\caption{\textbf{Baseline Result - Treated vs Control}: \footnotesize{In this figure, I plot the average yield for municipal bonds issued based on Equation \eqref{eq:dyn_yield_CompNego} in Panel (a). Panel (b) shows the differences between the yields of treated and control bonds. See Table \ref{table:vardescription} for variables description. The coefficients are shown in basis points. Specifically, the coefficients are obtained from regressing the yields on yearly interaction dummies for treated and control bonds, respectively, using issuer fixed effects. These coefficients are depicted on a yearly scale on the x-axis, where the vertical line corresponds to the Municipal Advisor Rule. The omitted benchmark period is the twelve-month period before the event window shown above. Standard errors are clustered by state. The dashed lines represent 95\% confidence intervals.}}
\label{fig:did_spread}
\end{figure}

\begin{figure}
    \raggedright\textbf{All Issuers}\\
    \begin{subfigure}{\linewidth}
    \centering
\includegraphics[scale=.75]{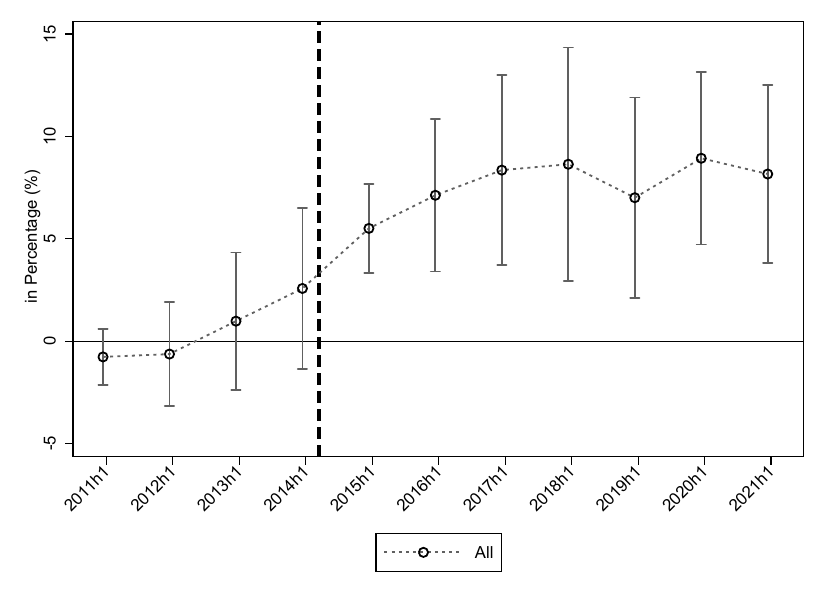}
    
    \caption{} 
    \label{fig:ll_advised}
    \end{subfigure}
    
    \medskip
    \raggedright\textbf{Large vs Small Issuers}\\
    \begin{subfigure}{\linewidth}
    \centering
\includegraphics[scale=.75]{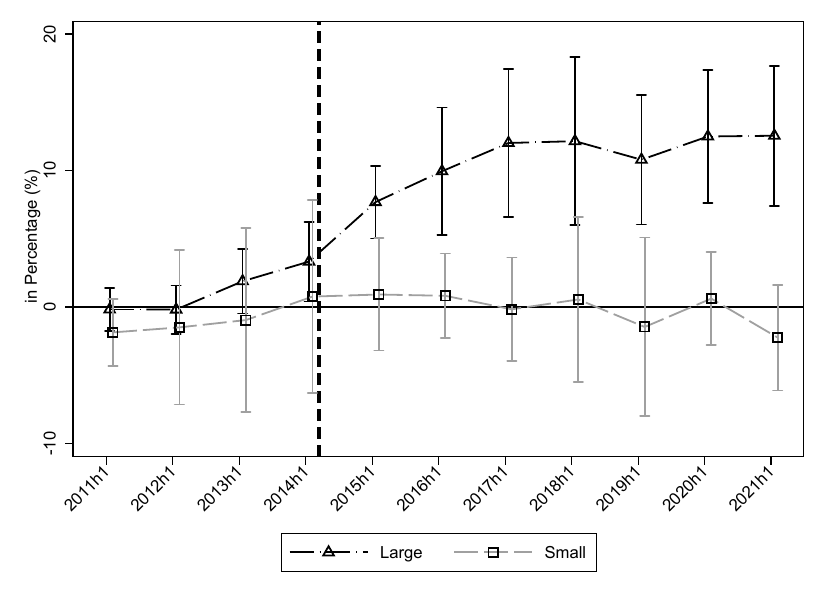}
    \caption{} 
    \label{fig:ll_advised_bySize}
    \end{subfigure}
\caption{\textbf{Likelihood of engaging advisor:} \footnotesize{This figure reports the coefficients showing the weighted average likelihood of issuing advised bonds on the y-axis, \textit{within} issuer. The vertical line corresponds to the SEC Municipal Advisor Rule. I represent these coefficients across all issuers in Panel (a), as well as the subsets of large versus small issuers in Panel (b). Standard errors are clustered at the state level. The dashed lines represent 95\% confidence intervals.}}
\label{fig:figure7}
\end{figure}

\begin{figure}
    \raggedright\textbf{New Municipal Bond Issuance}\\
    \begin{subfigure}{\linewidth}
    \centering
    \includegraphics[scale=0.75]{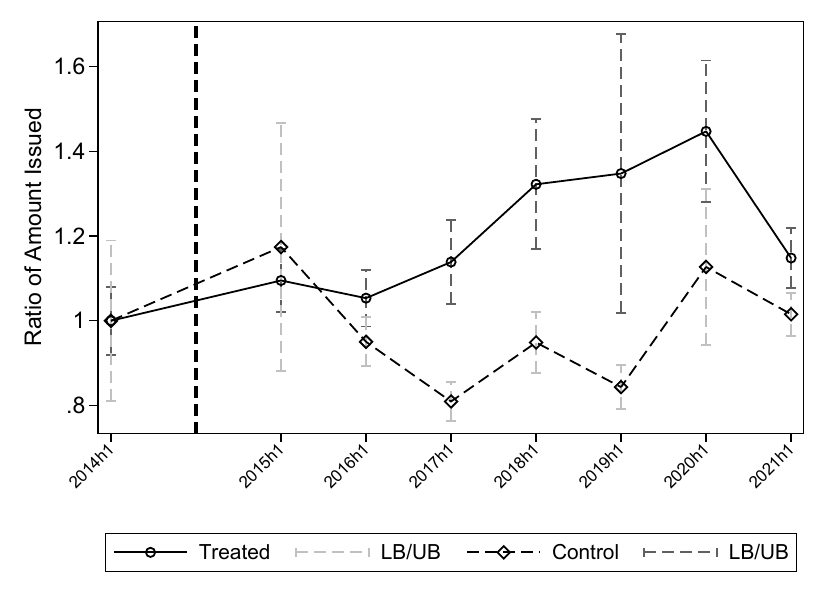}
    \caption{} 
    \label{fig:new_issuance_ov}
    \end{subfigure}
    
    \medskip
    \raggedright\textbf{New Municipal Bond Issuance by Size of Issuers}\\
    \begin{subfigure}{\linewidth}
    \centering
    \includegraphics[scale=0.85]{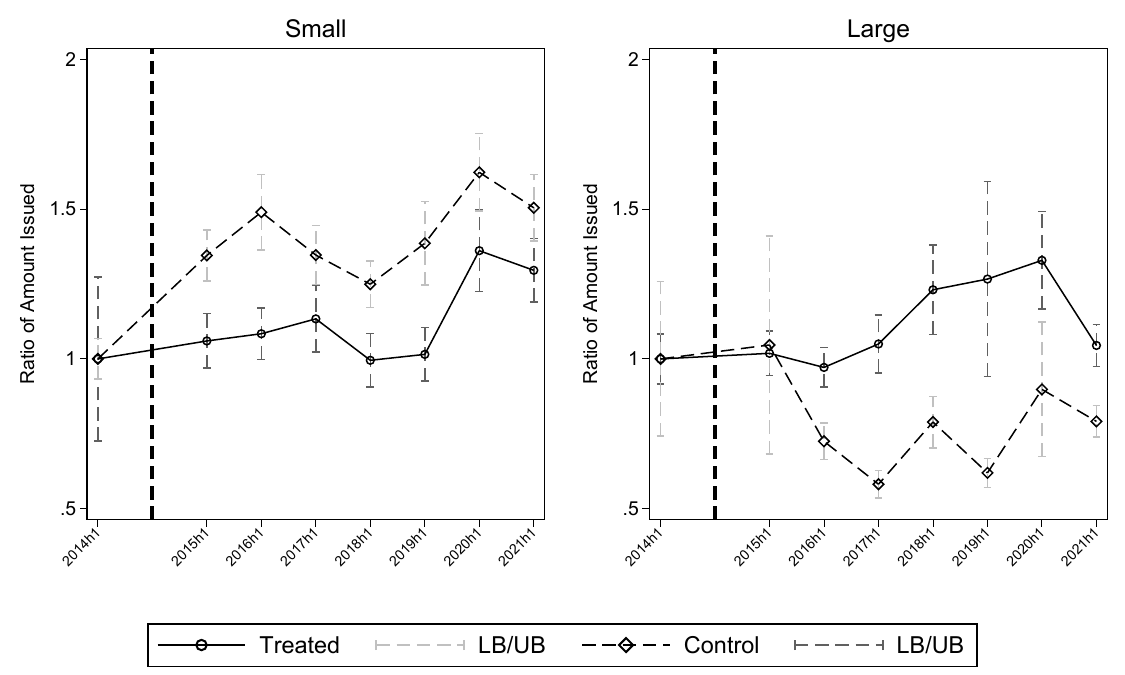}
    \caption{} 
    \label{fig:new_issuance_bySize}
    \end{subfigure}
\caption{\textbf{New Municipal Bond Issuance:} \footnotesize{In this figure, I plot the amount of municipal debt issued (with advisors) per capita in treated versus control issuers. The benchmark period is during the twelve months before the SEC Municipal Advisor Rule. Panel (a) shows results for all issuers. Panel (b) depicts the sub-samples corresponding to small versus large issuers. Standard errors are clustered by state. The dashed lines show the upper and lower limits based on the standard errors of the mean values.}}
\label{fig:figure8}
\end{figure}

\begin{figure}
\begin{center}
\textbf{Exit of Municipal Advising Firms}
\end{center}
\centering
\includegraphics[scale=1.25]{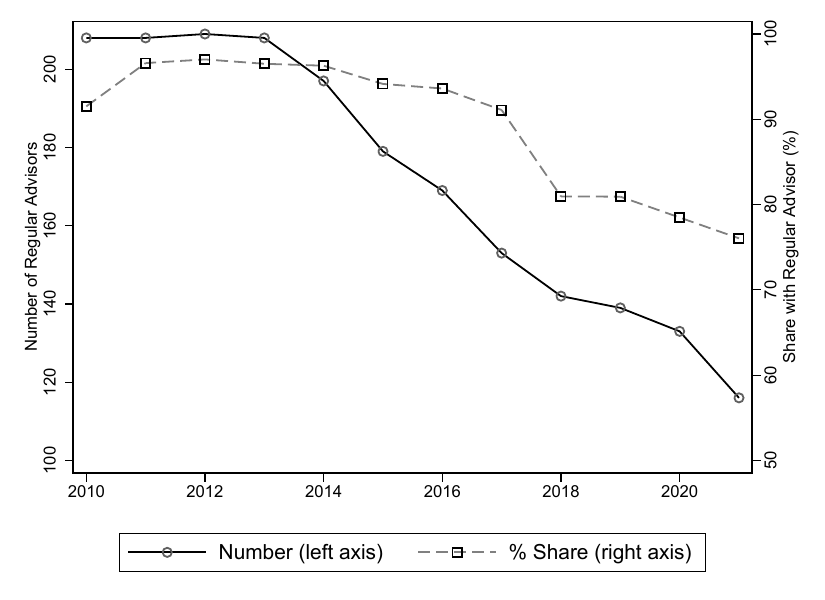}
\caption{\textbf{Number of Regular Advisors and Market Share:} \footnotesize{The figure shows the number of municipal advising firms on the left axis that regularly advise on municipal bond issuance before the SEC Municipal Advisor Rule. This corresponds to advisors in the sample who advise on at least one issuance in each calendar year until June 2014. On the right-hand axis, I plot the market share of these regular advisors during the sample period of 2010-2021. This represents the proportion of municipal debt advised by these advisors to the total municipal debt issued during the year, expressed as a percentage.}}
\label{fig:numAdv_Regular}
\end{figure}

\setcounter{table}{0}
\renewcommand{\thetable}{\arabic{table}}

\setcounter{figure}{0}
\renewcommand{\thefigure}{\arabic{figure}}

\clearpage

\begin{table}[htbp]
\def\sym#1{\ifmmode^{#1}\else\(^{#1}\)\fi}
\caption{Summary Statistics: Municipal Advisors and Municipal Bonds}
\label{table:summary_bonds1}
\vspace{0.2in}
\par
\footnotesize{This table shows the summary statistics for municipal advisors (MA) and municipal bonds in the sample. Panel A reports the top 15 municipal advisors in the data during the sample period of 2010-2021, who advise on fixed rate, tax-exempt bonds sold by municipalities. The number of issues corresponds to the aggregate number of bond issuances advised by the MA. The number of states in which MA's advise is split by total states versus the number of states across which over 50\% of the advisor's bond volume is spread. The share \% indicates the relative percentage of the advisor's volume in the sample. Panel B provides the municipal bond level characteristics during 2010-2021 for the bonds in the sample. The summary statistics correspond to the new issuance of bonds in the primary market. The key variables are described in Table \ref{table:vardescription}.}\vspace{.2in} \\

\centering
\raggedright\textbf{Panel A: Municipal Advisors}\\
\resizebox{1.0\textwidth}{!} {
\begin{tabular}{l*{6}{c}*{1}{H}}
\hline
                    &      Number of&      Average No. of &\multicolumn{2}{c}{Number of States}&  Total Volume&      Share(\%)&\multicolumn{1}{H}{Average}\\
                    \cline{4-5}\vspace{-1em}\\
                    &Issues&  Bonds Per Issue&\multicolumn{1}{c}{Total} &\multicolumn{1}{c}{$>$ 50\%ile}& (USD billion)&      &\multicolumn{1}{H}{Maturity (yrs.)}\\ 
\hline
THE PFM GROUP           &       8,937&          13&          50&           6&       574.3&        21.7&        10.1\\
PRAG&       1,233&          15&          29&           2&       347.6&        13.1&        11.1\\
FIRST SOUTHWEST     &       5,295&          15&          33&           1&       194.6&         7.4&        10.4\\
HILLTOP SECURITIES  &       1,907&          16&          28&           1&        83.0&         3.1&        10.7\\
ACACIA FINANCIAL&         917&          14&          14&           3&        69.1&         2.6&         9.7\\
MONTAGUE DEROSE &         307&          16&           5&           2&        58.9&         2.2&        11.6\\
LAMONT FINANCIAL &         304&          16&          12&           1&        53.7&         2.0&        11.1\\
RBC CAPITAL MARKETS &       1,662&          14&          13&           1&        52.9&         2.0&        10.2\\
KAUFMAN AND HALL    &         349&          12&          39&           7&        52.8&         2.0&        12.6\\
KNN PUBLIC FINANCE  &         603&          15&           1&           1&        47.4&         1.8&        11.0\\
PONDER \& COMPANY    &         343&          12&          38&           7&        43.7&         1.7&        12.5\\
PIPER JAFFRAY \& CO. &       1,612&          12&          20&           3&        35.2&         1.3&         9.8\\
DAVENPORT \& CO. LLC&         687&          17&          10&           2&        29.2&         1.1&        10.6\\
ESTRADA HINOJOSA&         636&          14&           2&           1&        28.7&         1.1&        10.5\\
FIELDMAN ROLAPP &         754&          16&           4&           1&        25.3&         1.0&        11.3\\
\hline
\end{tabular}
}

\vspace{4em}
\raggedright\textbf{Panel B: Municipal Bonds}\\
\centering
\resizebox{0.9\textwidth}{!} {
\begin{tabular}{l*{1}{cccccc}}
\hline
                    &\multicolumn{6}{c}{}                               \\
                    &       Count&        Mean&       Std. Dev.&   P25&  P50& P75         \\
\hline
Amount (USD million)&     937,994&        2.72&       11.63&         0.3&         0.6&         1.7\\
Coupon(\%)          &     937,994&        3.49&        1.15&         2.6&         3.4&         4.5\\
Years to Maturity   &     937,994&       10.13&        6.12&         5.2&         9.1&        14.0\\
Offering Price (USD)&     937,993&      107.24&        8.34&       100.0&       104.3&       112.8\\
Offering Yield(\%)  &     937,994&        2.25&        1.00&         1.5&         2.2&         3.0\\
Yield Spread(\%)    &     937,994&        1.22&        1.17&         0.4&         1.1&         2.0\\
Callable (Dummy)    &     937,994&        0.49&        0.50&         0.0&         0.0&         1.0\\
General Obligation (Dummy)&     937,994&        0.60&        0.49&         0.0&         1.0&         1.0\\
Bank Qualified (Dummy)&     937,994&        0.39&        0.49&         0.0&         0.0&         1.0\\
Cred. Enh. (Dummy)  &     937,994&        0.21&        0.41&         0.0&         0.0&         0.0\\
Insured (Dummy)     &     937,994&        0.17&        0.38&         0.0&         0.0&         0.0\\
\hline
\end{tabular}
}
\end{table}

\begin{table}[htbp]
\def\sym#1{\ifmmode^{#1}\else\(^{#1}\)\fi}
\caption{Impact on Offering Yield Spreads of Local Governments} 
\label{table:main_baseline}
\vspace{0.2in}
\par
{\footnotesize This table reports the baseline results for the sample using Equation \eqref{eq:baseline} estimating the differential effect on municipal bond yield spreads of treated and control bonds after the Municipal Advisor Rule of 2014. The primary coefficient of interest, $\beta_0$, is captured by the interaction term of \textit{Treated $\times$ Post}. I show the results using offering yield and offering yield spread as the dependent variable. I provide the description of key variables in Table \ref{table:vardescription}. In Table \ref{table:baseline_yld}, I show our results using offering yields as the dependent variable. T-statistics are reported in brackets and standard errors are clustered at the state level. \sym{*} \(p<0.10\), \sym{**} \(p<0.05\), \sym{***} \(p<0.01\)}\\
\vspace{.05in} 

\vspace{0.1in}
\centering
\resizebox{1.0\textwidth}{!} {
\begin{tabular}{@{\extracolsep{4pt}}l*{5}{c}}
\hline
                    \emph{Dependent Variable}:&\multicolumn{1}{c}{Offering Yield (bps)}&\multicolumn{4}{c}{Offering Yield Spread (bps)}\\
                    \cline{3-6}\\                    &\multicolumn{1}{c}{(1)}&\multicolumn{1}{c}{(2)}&\multicolumn{1}{c}{(3)}&\multicolumn{1}{c}{(4)}&\multicolumn{1}{c}{(5)}\\
\hline
Treated $\times$ Post &      -10.58\sym{**} &      -11.74\sym{***}&      -14.21\sym{***}&      -10.38\sym{***}&      -10.62\sym{***}\\
                    &     [-2.51]         &     [-2.86]         &     [-3.92]         &     [-2.96]         &     [-3.33]         \\
[1em]
Treated             &       -0.97         &        2.91         &       10.35\sym{***}&        8.46\sym{***}&        9.67\sym{***}\\
                    &     [-0.18]         &      [0.61]         &      [3.27]         &      [2.73]         &      [3.40]         \\
[1em]
Post                &      -14.77\sym{***}&      -11.12\sym{***}&       -8.17\sym{***}&       -9.16\sym{***}&        4.32\sym{**} \\
                    &     [-7.00]         &     [-5.73]         &     [-3.75]         &     [-4.49]         &      [2.33]         \\
[1em]
UW Mkt. Share(\%)   &                     &                     &                     &                     &       -7.28         \\
                    &                     &                     &                     &                     &     [-1.16]         \\
[1em]
$\triangle$Fed Funds$_{t-1}$&                     &                     &                     &                     &      -10.17\sym{***}\\
                    &                     &                     &                     &                     &     [-5.73]         \\
[1em]
$\triangle$UST10Y$_{t-1}$&                     &                     &                     &                     &       40.13\sym{***}\\
                    &                     &                     &                     &                     &     [31.74]         \\
[1em]
\hline
Issuer FE           &$\checkmark$         &$\checkmark$         &$\checkmark$         &$\checkmark$         &$\checkmark$         \\
State-Yr. FE        &$\checkmark$         &$\checkmark$         &$\checkmark$         &$\checkmark$         &$\checkmark$         \\
Bond Controls       &                     &                     &$\checkmark$         &$\checkmark$         &$\checkmark$         \\
Rating FE       &                     &                     &$\checkmark$         &$\checkmark$         &$\checkmark$         \\
County Controls     &                     &                     &$\checkmark$         &$\checkmark$         &$\checkmark$         \\
Adviser-Yr. FE      &                     &                     &                     &$\checkmark$         &$\checkmark$         \\
Adj.-R$^2$          &       0.323         &       0.648         &       0.826         &       0.837         &       0.850         \\
Obs.                &     937,845         &     937,845         &     937,845         &     937,819         &     937,819         \\
\hline
\end{tabular}
}
\end{table}

\begin{table}[htbp]
\def\sym#1{\ifmmode^{#1}\else\(^{#1}\)\fi}
\caption{Impact on Offering Price of New Bonds} 
\label{table:baseline_px}
\vspace{0.2in}
\par
{\footnotesize This table reports the baseline results for the sample using Equation \eqref{eq:baseline} estimating the differential effect on municipal bond offering prices of treated and control bonds after the Municipal Advisor Rule of 2014. The primary coefficient of interest, $\beta_0$, is captured by the interaction term of \textit{Treated $\times$ Post}. I show the results using offering price as the dependent variable. I provide the description of key variables in Table \ref{table:vardescription}. T-statistics are reported in brackets and standard errors are clustered at the state level. \sym{*} \(p<0.10\), \sym{**} \(p<0.05\), \sym{***} \(p<0.01\)}\\
\vspace{.05in} 

\vspace{0.1in}
\centering
\resizebox{0.85\textwidth}{!} {
\begin{tabular}{@{\extracolsep{4pt}}l*{4}{c}}
\hline
                    \emph{Dependent Variable}:&\multicolumn{4}{c}{Offering Price (per USD 100)}\\
                    \cline{2-5}\\               
                    &\multicolumn{1}{c}{(1)}&\multicolumn{1}{c}{(2)}&\multicolumn{1}{c}{(3)}&\multicolumn{1}{c}{(4)}\\
\hline
Treated $\times$ Post &        1.85\sym{***}&        1.37\sym{***}&        1.08\sym{***}&        1.10\sym{***}\\
                    &      [6.54]         &      [4.14]         &      [3.37]         &      [3.63]         \\
[1em]
Treated             &        0.79\sym{***}&       -0.46         &       -0.33         &       -0.38         \\
                    &      [3.58]         &     [-1.64]         &     [-1.32]         &     [-1.58]         \\
[1em]
Post                &        0.64\sym{***}&        0.68\sym{***}&        0.76\sym{***}&        0.16         \\
                    &      [3.12]         &      [3.63]         &      [4.64]         &      [1.01]         \\
[1em]
UW Mkt. Share(\%)   &                     &                     &                     &        0.01\sym{*}  \\
                    &                     &                     &                     &      [1.87]         \\
[1em]
$\triangle$Fed Funds$_{t-1}$&                     &                     &                     &        1.34\sym{***}\\
                    &                     &                     &                     &     [12.87]         \\
[1em]
$\triangle$UST10Y$_{t-1}$&                     &                     &                     &       -1.80\sym{***}\\
                    &                     &                     &                     &    [-21.77]         \\
[1em]

\hline
Issuer FE           &$\checkmark$         &$\checkmark$         &$\checkmark$         &$\checkmark$         \\
State-Yr. FE        &$\checkmark$         &$\checkmark$         &$\checkmark$         &$\checkmark$         \\
Bond Controls       &                     &$\checkmark$         &$\checkmark$         &$\checkmark$         \\
Rating FE       &                     &$\checkmark$         &$\checkmark$         &$\checkmark$         \\
County Controls     &                     &$\checkmark$         &$\checkmark$         &$\checkmark$         \\
Adviser-Yr. FE      &                     &                     &$\checkmark$         &$\checkmark$         \\
Adj.-R$^2$          &       0.407         &       0.785         &       0.793         &       0.798         \\
Obs.                &     937,844         &     937,844         &     937,818         &     937,818         \\
\hline
\end{tabular}
}
\end{table}

\begin{table}[htbp]
\def\sym#1{\ifmmode^{#1}\else\(^{#1}\)\fi}
\caption{Impact on Underwriter Spreads} 
\label{table:underpricing}
\vspace{0.2in}
\par
{\footnotesize This table reports the baseline results for the sample using Equation \eqref{eq:baseline} estimating the differential effect on municipal bond underwriting spreads of treated and control bonds after the Municipal Advisor Rule of 2014. The primary coefficient of interest, $\beta_0$, is captured by the interaction term of \textit{Treated $\times$ Post}. I show the results using underwriting spreads as the dependent variable. I provide the description of key variables in Table \ref{table:vardescription}. T-statistics are reported in brackets and standard errors are clustered at the state level. \sym{*} \(p<0.10\), \sym{**} \(p<0.05\), \sym{***} \(p<0.01\)}\\
\vspace{.05in} 

\vspace{0.1in}
\centering
\resizebox{0.85\textwidth}{!} {
\begin{tabular}{@{\extracolsep{4pt}}l*{4}{c}}
\hline
                    \emph{Dependent Variable}:&\multicolumn{4}{c}{Underwriting Spreads (basis points)}\\
                    \cline{2-5}\\               
                    &\multicolumn{1}{c}{(1)}&\multicolumn{1}{c}{(2)}&\multicolumn{1}{c}{(3)}&\multicolumn{1}{c}{(4)}\\
\hline
Treated $\times$ Post &      -13.22\sym{***}&      -12.48\sym{***}&      -13.21\sym{***}&      -13.30\sym{***}\\
                    &     [-7.15]         &     [-6.59]         &     [-8.65]         &     [-8.91]         \\
[1em]
Treated             &       -1.65         &        2.86         &        4.07\sym{**} &        4.24\sym{**} \\
                    &     [-0.72]         &      [1.28]         &      [2.41]         &      [2.50]         \\
[1em]
Post                &        0.32         &        1.21         &        1.30         &        3.23\sym{***}\\
                    &      [0.22]         &      [0.99]         &      [1.08]         &      [2.98]         \\
[1em]
UW Mkt. Share(\%)   &                     &                     &                     &        0.05         \\
                    &                     &                     &                     &      [0.54]         \\
[1em]
$\triangle$Fed Funds$_{t-1}$&                     &                     &                     &       -5.83\sym{***}\\
                    &                     &                     &                     &     [-6.81]         \\
[1em]
$\triangle$UST10Y$_{t-1}$&                     &                     &                     &        5.76\sym{***}\\
                    &                     &                     &                     &      [8.18]         \\
[1em]

\hline
Issuer FE           &$\checkmark$         &$\checkmark$         &$\checkmark$         &$\checkmark$         \\
State-Yr. FE        &$\checkmark$         &$\checkmark$         &$\checkmark$         &$\checkmark$         \\
Bond Controls       &                     &$\checkmark$         &$\checkmark$         &$\checkmark$         \\
Rating FE       &                     &$\checkmark$         &$\checkmark$         &$\checkmark$         \\
County Controls     &                     &$\checkmark$         &$\checkmark$         &$\checkmark$         \\
Adviser-Yr. FE      &                     &                     &$\checkmark$         &$\checkmark$         \\
Adj.-R$^2$          &       0.115         &       0.216         &       0.226         &       0.227         \\
Obs.                &     856,360         &     856,360         &     856,334         &     856,334         \\
\hline
\end{tabular}
}
\end{table}

\begin{table}[htbp]
\def\sym#1{\ifmmode^{#1}\else\(^{#1}\)\fi}
\caption{Impact on Liquidity of New Bonds} 
\label{table:baseline_disp}
\vspace{0.2in}
\par
{\footnotesize This table reports the baseline results for the sample using Equation \eqref{eq:baseline} estimating the differential effect on municipal bond liquidity of treated and control bonds after the Municipal Advisor Rule of 2014. The primary coefficient of interest, $\beta_0$, is captured by the interaction term of \textit{Treated $\times$ Post}. I show the results using price dispersion as the dependent variable. I provide the description of key variables in Table \ref{table:vardescription}. T-statistics are reported in brackets and standard errors are clustered at the state level. \sym{*} \(p<0.10\), \sym{**} \(p<0.05\), \sym{***} \(p<0.01\)}\\
\vspace{.05in} 

\vspace{0.1in}
\centering
\resizebox{0.85\textwidth}{!} {
\begin{tabular}{@{\extracolsep{4pt}}l*{4}{c}}
\hline
                    \emph{Dependent Variable}:&\multicolumn{4}{c}{Price Dispersion (per USD 100)}\\
                    \cline{2-5}\\               
                    &\multicolumn{1}{c}{(1)}&\multicolumn{1}{c}{(2)}&\multicolumn{1}{c}{(3)}&\multicolumn{1}{c}{(4)}\\
\hline
Treated $\times$ Post &       -0.03\sym{***}&       -0.03\sym{***}&       -0.03\sym{***}&       -0.03\sym{***}\\
                    &     [-6.35]         &     [-5.27]         &     [-5.58]         &     [-5.78]         \\
[1em]
Treated             &       -0.03\sym{***}&       -0.01         &       -0.01\sym{**} &       -0.01\sym{*}  \\
                    &     [-4.46]         &     [-1.54]         &     [-2.03]         &     [-1.88]         \\
[1em]
Post                &       -0.00         &        0.00         &        0.00         &        0.01\sym{*}  \\
                    &     [-0.39]         &      [0.50]         &      [0.03]         &      [1.85]         \\
[1em]
UW Mkt. Share(\%)   &                     &                     &                     &        0.00\sym{***}\\
                    &                     &                     &                     &      [3.50]         \\
[1em]
$\triangle$Fed Funds$_{t-1}$&                     &                     &                     &       -0.02\sym{***}\\
                    &                     &                     &                     &     [-6.62]         \\
[1em]
$\triangle$UST10Y$_{t-1}$&                     &                     &                     &        0.02\sym{***}\\
                    &                     &                     &                     &     [25.51]         \\
[1em]
\hline
Issuer FE           &$\checkmark$         &$\checkmark$         &$\checkmark$         &$\checkmark$         \\
State-Yr. FE        &$\checkmark$         &$\checkmark$         &$\checkmark$         &$\checkmark$         \\
Bond Controls       &                     &$\checkmark$         &$\checkmark$         &$\checkmark$         \\
Rating FE       &                     &$\checkmark$         &$\checkmark$         &$\checkmark$         \\
County Controls     &                     &$\checkmark$         &$\checkmark$         &$\checkmark$         \\
Adviser-Yr. FE      &                     &                     &$\checkmark$         &$\checkmark$         \\
Adj.-R$^2$          &       0.128         &       0.305         &       0.313         &       0.314         \\
Obs.                &     856,361         &     856,361         &     856,335         &     856,335         \\
\hline
\end{tabular}
}
\end{table}

\begin{table}[htbp]
\def\sym#1{\ifmmode^{#1}\else\(^{#1}\)\fi}
\caption{Impact Due to Advisor's Ex-ante Role in ``Selecting'' Underwriter} 
\label{table:cs_byUWintro}
\vspace{0.2in}
\par
{\footnotesize This table reports the results using Equation \eqref{eq:baseline} to show the differential effect among issuers for whom advisors play a greater role in selecting underwriters. The dependent variable is offering yield spread. I use the average (Columns (1)-(3)) and weighted average (Columns (4)-(6)) of ex-ante likelihood of a new underwriter being introduced by an advisor for a given issuer, respectively. Specifically, I interact the equation with dummies corresponding to below and above median values for this measure among issuers. This analysis also includes group $\times$ year fixed effects. The baseline specification of Column (5) in Table \ref{table:main_baseline} is shown in Columns (3) and (6). T-statistics are reported in brackets and standard errors are clustered at the state level. \sym{*} \(p<0.10\), \sym{**} \(p<0.05\), \sym{***} \(p<0.01\)}\\
\vspace{.05in} 

\vspace{0.1in}
\centering
\resizebox{1.0\textwidth}{!} {
\begin{tabular}{@{\extracolsep{4pt}}l*{6}{c}}
\hline
                    \emph{Dependent Variable}:&\multicolumn{6}{c}{Yield Spread (basis points)}\\
                    \cline{2-7}\\ 
                    \emph{Based on Issuers'}:&\multicolumn{3}{c}{Average}&\multicolumn{3}{c}{Weighted Average}\\
                    \cline{2-4}\cline{5-7}\\ 
    Treated $\times$ Post&\multicolumn{1}{c}{(1)}&\multicolumn{1}{c}{(2)}&\multicolumn{1}{c}{(3)}&\multicolumn{1}{c}{(4)}&\multicolumn{1}{c}{(5)}&\multicolumn{1}{c}{(6)}\\
\hline
$\times$ Below Median       &      -10.03         &       -3.14         &       -0.73         &      -10.03         &       -3.14         &       -0.73         \\
                    &     [-1.47]         &     [-0.52]         &     [-0.13]         &     [-1.47]         &     [-0.52]         &     [-0.13]         \\
[1em]
$\times$ Above Median       &      -19.01\sym{***}&      -16.48\sym{***}&      -13.55\sym{***}&      -19.01\sym{***}&      -16.48\sym{***}&      -13.55\sym{***}\\
                    &     [-7.21]         &     [-7.61]         &     [-6.03]         &     [-7.22]         &     [-7.61]         &     [-6.03]         \\
[1em]
\hline
Difference          &        8.99         &       13.35         &       12.83         &        8.99         &       13.35         &       12.83         \\
p-value             &        0.12         &        0.02         &        0.01         &        0.12         &        0.02         &        0.01         \\
Issuer FE           &$\checkmark$         &$\checkmark$         &$\checkmark$         &$\checkmark$         &$\checkmark$         &$\checkmark$         \\
Controls            &$\checkmark$         &$\checkmark$         &$\checkmark$         &$\checkmark$         &$\checkmark$         &$\checkmark$         \\
Advisor FE          &$\checkmark$         &$\checkmark$         &         &$\checkmark$         &$\checkmark$         &         \\
Group-Yr. FE        &$\checkmark$         &$\checkmark$         &$\checkmark$         &$\checkmark$         &$\checkmark$         &$\checkmark$         \\
State-Yr. FE        &                     &$\checkmark$         &$\checkmark$         &                     &$\checkmark$         &$\checkmark$         \\
Advisor-Yr. FE      &                     &                     &$\checkmark$         &                     &                     &$\checkmark$         \\
Adj.-R$^2$          &       0.835         &       0.840         &       0.848         &       0.835         &       0.840         &       0.848         \\
Obs.                &     859,230         &     859,230         &     859,210         &     859,230         &     859,230         &     859,210         \\
\hline
\end{tabular}
}
\end{table}

\begin{table}[htbp]
\def\sym#1{\ifmmode^{#1}\else\(^{#1}\)\fi}
\caption{Heterogeneity by Size of Issuers} 
\label{table:cs_bySize}
\vspace{0.2in}
\par
{\footnotesize This table reports the results using Equation \eqref{eq:baseline} to show the differential effect among issuers based on their ex-ante size. I use the average (Columns (1)-(3)) and median (Columns (4)-(6)) size of ex-ante issuances, respectively. Specifically, I interact the equation with dummies corresponding to small and large values for this measure among issuers. This analysis also includes group $\times$ year fixed effects. The baseline specification of Column (5) in Table \ref{table:main_baseline} is shown in Columns (3) and (6). T-statistics are reported in brackets and standard errors are clustered at the state level. \sym{*} \(p<0.10\), \sym{**} \(p<0.05\), \sym{***} \(p<0.01\)}\\
\vspace{.05in} 

\vspace{0.1in}
\centering
\resizebox{1.0\textwidth}{!} {
\begin{tabular}{@{\extracolsep{4pt}}l*{6}{c}}
\hline
                    \emph{Dependent Variable}:&\multicolumn{6}{c}{Yield Spread (basis points)}\\
                    \cline{2-7}\\ 
                    \emph{Based on Issuers'}:&\multicolumn{3}{c}{Average size, ex-ante}&\multicolumn{3}{c}{Median size, ex-ante}\\
                    \cline{2-4}\cline{5-7}\\ 
    Treated $\times$ Post&\multicolumn{1}{c}{(1)}&\multicolumn{1}{c}{(2)}&\multicolumn{1}{c}{(3)}&\multicolumn{1}{c}{(4)}&\multicolumn{1}{c}{(5)}&\multicolumn{1}{c}{(6)}\\
\hline
$\times$ Small&       -5.86         &        0.13         &        0.56         &       -3.64         &        1.68         &        1.72         \\
                    &     [-0.66]         &      [0.02]         &      [0.07]         &     [-0.40]         &      [0.20]         &      [0.20]         \\
[1em]
$\times$ Large&      -19.73\sym{***}&      -15.90\sym{***}&      -12.60\sym{***}&      -20.65\sym{***}&      -16.96\sym{***}&      -13.58\sym{***}\\
                    &     [-7.10]         &     [-7.36]         &     [-5.60]         &     [-7.22]         &     [-7.39]         &     [-5.91]         \\
[1em]
\hline
Difference          &       13.87         &       16.03         &       13.16         &       17.00         &       18.64         &       15.29         \\
p-value             &        0.09         &        0.03         &        0.08         &        0.07         &        0.04         &        0.09         \\
Issuer FE           &$\checkmark$         &$\checkmark$         &$\checkmark$         &$\checkmark$         &$\checkmark$         &$\checkmark$         \\
Controls            &$\checkmark$         &$\checkmark$         &$\checkmark$         &$\checkmark$         &$\checkmark$         &$\checkmark$         \\
Advisor FE          &$\checkmark$         &$\checkmark$         &         &$\checkmark$         &$\checkmark$         &         \\
Group-Yr. FE        &$\checkmark$         &$\checkmark$         &$\checkmark$         &$\checkmark$         &$\checkmark$         &$\checkmark$         \\
State-Yr. FE        &                     &$\checkmark$         &$\checkmark$         &                     &$\checkmark$         &$\checkmark$         \\
Advisor-Yr. FE      &                     &                     &$\checkmark$         &                     &                     &$\checkmark$         \\
Adj.-R$^2$          &       0.838         &       0.842         &       0.850         &       0.838         &       0.842         &       0.850         \\
Obs.                &     937,840         &     937,840         &     937,819         &     937,840         &     937,840         &     937,819         \\
\hline
\end{tabular}
}
\end{table}

\begin{table}[htbp]
\def\sym#1{\ifmmode^{#1}\else\(^{#1}\)\fi}
\caption{Heterogeneity by Sophistication of Issuers} 
\label{table:cs_bySophis}
\vspace{0.2in}
\par
{\footnotesize This table reports the results using Equation \eqref{eq:baseline} to show the differential effect among issuers based on their ex-ante size. The dependent variable is offering yield spread. I use the average (Columns (1)-(3)) and median (Columns (4)-(6)) size of ex-ante issuances, respectively. Specifically, I interact the equation with dummies corresponding to small and large values for this measure among issuers. This analysis also includes group $\times$ year fixed effects. The baseline specification of Column (5) in Table \ref{table:main_baseline} is shown in Columns (3) and (6). T-statistics are reported in brackets and standard errors are clustered at the state level. \sym{*} \(p<0.10\), \sym{**} \(p<0.05\), \sym{***} \(p<0.01\)}\\
\vspace{.05in} 

\vspace{0.1in}
\centering
\resizebox{1.0\textwidth}{!} {
\begin{tabular}{@{\extracolsep{4pt}}l*{5}{c}}
\hline
                    \emph{Dependent Variable}:&\multicolumn{5}{c}{Yield Spread (basis points)}\\
                    \cline{2-6}\\ 
                    \emph{Based on Issuers'}:&\multicolumn{2}{c}{Complexity of Bonds (ex-ante)}&\multicolumn{1}{c}{Credit}&\multicolumn{1}{c}{Average Wages}&\multicolumn{1}{c}{Fraction}\\
                    \cline{2-3}\\
                    &\multicolumn{1}{c}{All}&\multicolumn{1}{c}{Advised}&\multicolumn{1}{c}{Enhancement}&\multicolumn{1}{c}{of Finance Staff}&\multicolumn{1}{c}{advised}\\
                    \cline{2-3}\\
    Treated $\times$ Post&\multicolumn{1}{c}{(1)}&\multicolumn{1}{c}{(2)}&\multicolumn{1}{c}{(3)}&\multicolumn{1}{c}{(4)}&\multicolumn{1}{c}{(5)}\\
\hline
$\times$ Below Median&       -3.92         &       -6.26\sym{***}&       -7.70\sym{**} &       -5.14         &                     \\
                    &     [-1.46]         &     [-2.84]         &     [-2.16]         &     [-1.55]         &                     \\
[1em]
$\times$ Above Median&      -16.84\sym{***}&      -13.61\sym{***}&      -14.75\sym{***}&      -12.66\sym{***}&                     \\
                    &     [-6.20]         &     [-3.80]         &     [-6.16]         &     [-5.38]         &                     \\
[1em]
$\times$ High&                     &                     &                     &                     &      -10.36\sym{***}\\
                    &                     &                     &                     &                     &     [-3.23]         \\
[1em]
$\times$ Low&                     &                     &                     &                     &      -13.81\sym{***}\\
                    &                     &                     &                     &                     &     [-5.36]         \\
[1em]
\hline
Difference          &       12.92         &        7.35         &        7.05         &        7.51         &       3.45         \\
p-value             &        0.00         &        0.01         &        0.02         &        0.04         &        0.30         \\
Issuer FE           &$\checkmark$         &$\checkmark$         &$\checkmark$         &$\checkmark$         &$\checkmark$         \\
Advisor-Year FE          &$\checkmark$         &$\checkmark$         &$\checkmark$         &$\checkmark$         &$\checkmark$         \\
Controls       &$\checkmark$         &$\checkmark$         &$\checkmark$         &$\checkmark$         &$\checkmark$         \\
State-Year FE       &$\checkmark$         &$\checkmark$         &$\checkmark$         &$\checkmark$         &$\checkmark$         \\
Group-Year FE       &$\checkmark$         &$\checkmark$         &$\checkmark$         &$\checkmark$         &$\checkmark$         \\
Adj.-R$^2$          &       0.848         &       0.847         &       0.848         &       0.850         &       0.848         \\
Obs.                &     805,329         &     769,976         &     801,894         &     687,432         &     801,919         \\
\hline
\end{tabular}
}
\end{table}

\begin{table}[htbp]
\def\sym#1{\ifmmode^{#1}\else\(^{#1}\)\fi}
\caption{Evidence from States' Response to FOIA Requests} 
\label{table:foia_st}
\vspace{0.2in}
\par
{\footnotesize This table reports the results using Equation \eqref{eq:baseline} estimating the differential effect on yield spreads of treated and control bonds after the Municipal Advisor Rule of 2014 between states that did and did not respond with data on municipal advisor fees for the FOIA requests. In Columns (1)-(2), I show the results using offering yield spreads as the dependent variable using sub-samples of states. I restrict the sub-sample to issuers in states that responded with fee data to FOIA requests (FOIA Data $=$ Yes) in Column (1). In Column (2), I restrict the sub-sample to issuers in states that did not respond with fee data to FOIA requests (FOIA Data $=$ No) in Column (2). The primary coefficient of interest, $\beta_0$, is captured by the interaction term of \textit{Treated $\times$ Post}. Finally, Column (3) reports the results for the full sample using an interacted specification. T-statistics are reported in brackets and standard errors are clustered at the state level. \sym{*} \(p<0.10\), \sym{**} \(p<0.05\), \sym{***} \(p<0.01\)}\\
\vspace{.05in} 

\vspace{0.1in}
\centering
\resizebox{0.9\textwidth}{!} {
\begin{tabular}{@{\extracolsep{4pt}}l*{3}{c}}
\hline
                    \emph{Dependent Variable}:&\multicolumn{3}{c}{Yield Spread (basis points)}\\
                    \cline{2-4} \\
                    \emph{Sample of Issuers}:&\multicolumn{2}{c}{Sub-samples of States}&\multicolumn{1}{c}{All}\\
                    \cline{2-3} \cline{4-4} \\
                    &\multicolumn{1}{c}{FOIA Data=Yes}&\multicolumn{1}{c}{FOIA Data=No}\\                    

Treated $\times$ Post&\multicolumn{1}{c}{(1)}&\multicolumn{1}{c}{(2)}&\multicolumn{1}{c}{(3)}\\
\hline
 &       -3.77         &      -17.85\sym{***}&                     \\
                    &     [-1.15]         &     [-5.99]         &                     \\
[1em]
$\times$ $\mathds{1}$ (FOIA Data=Yes)&                     &                     &       -4.59         \\
                    &                     &                     &     [-1.29]         \\
[1em]
$\times$ $\mathds{1}$ (FOIA Data=No) &                     &                     &      -17.68\sym{***}\\
                    &                     &                     &     [-5.81]         \\
[1em]
\hline
Difference          &                     &                     &       13.09         \\
p-value             &                     &                     &        0.01         \\
Issuer FE           &$\checkmark$         &$\checkmark$         &$\checkmark$         \\
State-Yr. FE        &$\checkmark$         &$\checkmark$         &$\checkmark$         \\
Controls       &$\checkmark$  &$\checkmark$         &$\checkmark$         \\
Group-Yr. FE          &         &         &$\checkmark$         \\
Advisor-Yr. FE          &$\checkmark$         &$\checkmark$         &$\checkmark$\\
Adj.-R$^2$          &       0.847         &       0.855         &       0.850         \\
Obs.                &     452,235         &     485,580         &     937,819         \\

\hline
\end{tabular}
}
\end{table}

\begin{table}[htbp]
\def\sym#1{\ifmmode^{#1}\else\(^{#1}\)\fi}
\caption{Evidence from the Exit of Municipal Advisors (MA)} 
\label{table:exitingMA_over50}
\vspace{0.2in}
\par
{\footnotesize This table reports the results using Equation \eqref{eq:baseline} with interactions to show the differential effect among issuers based on the exit of municipal advisors (MA). This analysis also includes group $\times$ year fixed effects. Column (1) shows results among all issuers with interactions corresponding to whether the issuer primarily depended on an exiting  advisor or not. I define issuers linked to advisors when more than 50\% of their municipal debt issuance in the pre-period is advised by the exiting advisor. I focus on the exit of regular advisors. These represent municipal advisors with at least one issuance in each calendar year before the SEC Municipal Advisor Rule in the sample. Columns (2) and (3) show results for issuers that depend on exiting advisors. For these issuers, I show the heterogeneity between small and large issuers based on the median size of ex-ante issuances. Column (2) shows results for advised bonds only. Column (3) also includes bonds issued without any advisors, and the analysis does not include advisor $\times$ year fixed effects. T-statistics are reported in brackets and standard errors are clustered at the state level. \sym{*} \(p<0.10\), \sym{**} \(p<0.05\), \sym{***} \(p<0.01\)}\\
\vspace{.05in} 

\vspace{0.1in}
\centering
\resizebox{0.9\textwidth}{!} {
\begin{tabular}{@{\extracolsep{4pt}}l*{3}{c}}
\hline
                    \emph{Dependent Variable}:&\multicolumn{3}{c}{Yield Spread (basis points)}\\
                    \cline{2-4} \\
                    \emph{Sample of Issuers}:&\multicolumn{1}{c}{All}&\multicolumn{2}{c}{Dependent on Exiting MA}\\
                    \cline{2-2} \cline{3-4} \\
Treated $\times$ Post&\multicolumn{1}{c}{(1)}&\multicolumn{1}{c}{(2)}&\multicolumn{1}{c}{(3)}\\
\hline
$\times$ Other   &      -13.02\sym{***}&                     &                     \\
                    &     [-4.16]         &                     &                     \\
[1em]
$\times$ Dependent on Exiting MA     &       -3.21         &                     &                     \\
                    &     [-0.79]         &                     &                     \\
[1em]
$\times$ Small      &                     &       15.78\sym{***}&       14.38\sym{*}  \\
                    &                     &      [3.32]         &      [1.95]         \\
[1em]
$\times$ Large      &                     &      -12.09\sym{***}&      -10.95\sym{***}\\
                    &                     &     [-3.09]         &     [-3.07]         \\
[1em]
\hline
Difference          &        -9.82         &       27.88         &       25.32         \\
p-value             &        0.03         &        0.00         &        0.00         \\
Issuer FE           &$\checkmark$         &$\checkmark$         &$\checkmark$         \\
State-Yr. FE        &$\checkmark$         &$\checkmark$         &$\checkmark$         \\
Controls       &$\checkmark$  &$\checkmark$         &$\checkmark$         \\
Group-Yr. FE          &$\checkmark$         &$\checkmark$         &$\checkmark$         \\
Advisor-Yr. FE          &$\checkmark$         &$\checkmark$         &         \\
Adj.-R$^2$          &       0.850         &       0.865         &       0.853         \\
Obs.                &     937,819         &     250,623         &     250,632         \\
\hline
\end{tabular}
}
\end{table}

\clearpage
\newpage
\begin{itemize}\centering
	\item [] {\bf \large For Online Publication--Internet Appendix}\\ \vspace{0.2cm}
\end{itemize}

\thispagestyle{empty}
\appendix

\begin{appendices}


\setcounter{table}{0}
\setcounter{page}{0}
\renewcommand{\thepage}{IA\arabic{page}}
\renewcommand{\thetable}{IA\arabic{table}}

\setcounter{figure}{0}
\renewcommand{\thefigure}{IA\arabic{figure}}

\renewcommand{\thesection}{IA\arabic{section}}
\setcounter{section}{0}
\setcounter{equation}{0}

\clearpage
\doublespacing

\section{SEC Municipal Advisor Rule}\label{sec:sec_ma_rule}
As shown in Figure \ref{fig:total_iss}, municipalities issue over USD 400 billion of municipal bonds each year to finance various infrastructure and public utility projects. However, these municipalities may often lack the financial sophistication to navigate the issuance process \citep{garrett2021conflicts}. Under the Congressional mandate of June 1975, the MSRB has been charged with protecting investors to prevent fraud and financial irregularities in the municpial bond market\footnote{\href{https://www.msrb.org/About-MSRB/About-the-MSRB/Creation-of-the-MSRB.aspx}{https://www.msrb.org/About-MSRB/About-the-MSRB/Creation-of-the-MSRB.aspx}}. Following the financial crisis of 2007-09, Congress passed the Dodd-Frank Wall Street Reform and Consumer Protection Act in 2010. This Act introduced major changes to the regulatory framework and operations of the financial service industry. It also included several provisions concerning SEC rulemaking.

Under this framework, the SEC drew up the Municipal Advisor Rule which became effective on July 1, 2014. Specifically, it became unlawful for a municipal advisor (MA) to provide advice to or on behalf of a municipal entity or obligated person with respect to municipal financial products or the issuance of municipal securities unless the MA is registered with the SEC. Prior to this reform, only broker-dealers and banks were subject to federal regulatory requirements. The SEC sought to mitigate some of the problems observed in the conduct of municipal advisors in the municipal bond market. This included the municipal advisor's failure to place the duty of loyalty to their clients ahead of their own interests \citep{white2014}. Additionally, the SEC Commissioner noted in a statement in 2013\footnote{\href{https://www.sec.gov/news/statement/2013-09-18-open-meeting-statement-kms}{https://www.sec.gov/news/statement/2013-09-18-open-meeting-statement-kms}}:
\vspace{-1em}
\begin{verse}
\noindent
\textit{Our dedicated public servants were relying on municipal advisors whose advisory activities generally did not require them to register with the Commission, or any other federal, state, or self-regulatory entity.    And a lack of meaningful regulation over these advisors created confusion, and in some instances, horrific abuses.  Sadly, the shortcomings of this hands-off regulatory regime became glaringly apparent during the last several years as we learned about numerous examples of bad behavior, including self-dealing and excessive fees.}
\end{verse}

Importantly, advisors owe fiduciary responsibility to municipal clients under the MA Rule and cannot act in ways that my be unfavorable to their clients. This would encompass the twin requirements of duty of care, and duty of loyalty. According to the former, advisors must exert effort on behalf of the issuer in order to make a recommendation. The duty of loyalty requires advisors to uphold the interests of the issuer superior to their own. Failing to adhere to their fiduciary responsibility, advisors may be held liable for adverse outcomes to issuers during municipal bond issuance. The SEC MA Rule also clarified what constitutes municipal advice and therefore under what circumstances the registration requirements would be applicable. Broadly, the SEC documented ``advice'' as any recommendation particularized to specific needs, objectives, and circumstances of a municipal entity. Overall, the SEC Municipal Advisor Rule introduced a new set of standards applicable to municipal advisors, with an aim to help municipal issuers.

\section{Advisor Names}\label{sec:adv_names}
There is substantial variation in the names used by FTSE Russell municipal bond database to record municipal advisors involved in the issuance of bonds. Often these variations are due to spellings like ``BACKSTORM MCCARLEY BERRY \& COMPANY LLC'' vs ``BACKSTROM MCCARLEY BERRY \& COMPANY''. I investigate the bond offer statement associated with the CUSIP to verify the corresponding entity and update the standardized name. Besides the jumbled spelling errors, I also come across typos and mistakes due to omitted letters. For example, ``BERKSHIRE BANK'' and ``BERSHIRE BANK'' correspond to the same municipal advisor. Further, I also account for alternative company name extensions, such as, ``LLC'', ``INC'', ``Co.'' or ``CO.''. These extensions are recorded differently over time for the same company. Occasionally, the names would omit portions of the names altogether: ``SUDSINA \& ASSOCIATES'' was also reported as ``SUDSINA''. To verify against mismatching firm names, I rely on logos printed on the bond's offering statement in addition to the official address provided. 

Other instances of spelling differences involve special characters like ``.'' or ``,'' and ``\&''. Clearly, these are easy to handle and to resolve. I also need to update names for subsidiaries and affiliated company names. These would also often involve mergers and acquisitions. Following \cite{cestau2020specialization}, I retroactively replace the names of the merged entities with the name of the new company. To an extent, this assumes that client relationships prevailed even after the M\&A activity. Indeed I do find evidence where both sets of names are found in the offering statements. This could also be due to delay in sale of bonds, especially via negotiation. Thus, offer statements may reflect contractual engagements before or after the merger activity. As a result, I identify these entities under a common name: for example, ``MERRILL LYNCH \& COMPANY'' and ``BANC OF AMERICA SECURITIES'' as ``BANC OF AMERICA SECURITIES''. I also find limited anecdotal evidence suggesting that acquisitions involve retention of officials in the new company.

Still other challenges include names where firms operate using different brand names. For example, ``SUSAN D. MUSSELMAN INC'', ``SDM ADVISORS INC'' and ``DASHENMUSSELMAN INC'' are captured differently. However, I represent them as ``A DASHEN \& ASSOCIATES'' by checking the names of principals and office locations recorded under each company. I believe that it is very difficult to identify abbreviated names corresponding to the same umbrella company without verifying each name separately. In some cases, I need to ascertain the office locations from internet search versus that reported in the bond offering statement. But I also come across simpler associations where alternative trading names used by companies are seemingly related, such as, ``RV NORENE \& ASSOCIATE INC'' and ``CROWE NORENE''.

\section{\label{subsec:robustness}Robustness}
In this section, I test the robustness of the main result in Column (5) of Table \ref{table:main_baseline} to various alternative econometric considerations. I present the results of these robustness checks in Table \ref{table:robustness} ranging across alternative specifications, tax and bond considerations,  geographic considerations as well as alternative clustering of standard errors.

\subsection{Alternative Specifications}\label{subsubsec:alt_spec}\noindent
In Column (1), I show the baseline results by adding issuer-type $\times$ year fixed effects. By augmenting the model, I add flexible time trends for different types of issuers (city, county, state or other) interacted with year fixed effects. Next, I show results after controlling for the underlying purpose of bonds in Column (2). I also control for such unobserved trends over time by including bond purpose $\times$ year fixed effects in Column (3). I find that the magnitude (-10.08 bps) is nearly similar to the main effect and is statistically significant. Thus, I rule out explanations about the main effect that may be linked to trends in the purpose of bonds which may change around the same time as the imposition of fiduciary duty.

\citet*{cestau2020specialization} shows that underwriters tend to specialize in the method of sale in the municipal bond market. Therefore, I address the possibility that unobserved changes among underwriters may simultaneously affect yields as the SEC Municipal Advisor Rule. First, in Column (4), I include underwriter fixed effects to the main specification to absorb unobserved characteristics across underwriters. Thereafter, I also absorb unobserved time-varying trends associated with underwriters (underwriter $\times$ year fixed effects) and present the results in Column (5). The reported coefficient is -9.26 bps. Overall, I argue that the results are robust to unobserved time-unvarying and time-varying changes related to underwriters.

There may be a concern that issuers and advisors may rely on past relationships during new municipal bond issuances, or that advisors specialize in a given state. I account for this unobserved effect by including issuer-advisor pair fixed effect in Column (6), and find that the main result is robust to this consideration. I also introduce issuer-underwriter pair fixed effect to the main specification in Column (7). The result shows that the estimate is similar (-14.02 bps) after accounting for this. I examine the sensitivity of the baseline result to controlling for unobserved advisor-state pairing. Column (8) shows that the baseline effect is robust to this consideration. Finally, I also show results for a restrictive specification in Column (9), by including county $\times$ year fixed effect. Offering yield spreads may be unobservably driven by local economic conditions at the county level. Even with this granular fixed effect, I find that the main coefficient is only slightly lower (-8.79 bps) and remains statistically significant.

\subsection{Additional Considerations on Taxability, Bond features and Geography}\noindent
Given the heterogeneous effects due to tax considerations, I show robustness of the main results to these aspects in Columns (10)-(12). First, in Column (10), I broaden the sample to bonds for which interest income is taxable under federal law. I find the effect to be -10.62 bps and statistically significant. Next, I drop bonds from states that do not provide income tax exemption for in-state or out-of-state municipal bonds (IL, IA, KS, OK, WI)  \citep*{gao2021good}. Here, I report a baseline estimate of -9.27 bps in Column (11). Finally, I only focus on bonds that are exempt from federal as well as state level taxes in Column (12). The reported coefficient is -10.27 bps and is statistically significant. Thus, I conclude that the results are not sensitive to these tax considerations on interest income from municipal bonds.

For the baseline analysis, I present the results using a wide variety of municipal bonds. For example, bonds that are advised and underwritten by the same financial agent may be unobservably different and may confound the estimates. However, in Column (13) of Table \ref{table:robustness}, I show the results by dropping a small number of such bonds during the sample period and find a similar effect. Next, to the extent that yields may vary unobservably differently for callable bonds around the same time as the SEC regulation, the estimates may be confounded. Therefore, I show the baseline effect by dropping callable bonds in Column (14) and show that the findings are robust. With similar considerations, I also show results by dropping insured bonds (in Column (15)) and keeping only new money bonds (in Column (16)), respectively. I find that the baseline magnitude increases for these sub-samples and remains statistically significant.

Finally, I turn to geographic considerations that may confound with the identification strategy. First, local bonds may be different from state level bonds/issuers. As a result, I show the result by keeping only local bonds in Column (17), followed by restricting the sample to state level bonds only in Column (18). While the magnitude reduces marginally to -9.29 bps in the former, I find a greater impact (-12.85 bps) for state level bonds. The greater effect on state bonds is consistent with the higher impact on more sophisticated issuers, discussed in Section \ref{subsec:hetero_size}. Next, I also show robustness of the results by dropping observations from the largest states (California, New York and Texas) in Column (19) to show that the effect is not driven by these states alone. 

\subsection{Alternative Clustering of Standard Errors}\noindent
I follow a conservative approach in clustering standard errors by state in the baseline specification. However, I consider alternative levels of clustering standard errors in Columns (20)-(30). First, I think of modifying the cross-sectional dimension of observations. In Column (20), I cluster standard errors by advisor and find the main result holds. The baseline effect is also robust to clustering standard errors by underwriters (Column (21)). Column (22) shows results by clustering standard errors by issuer. I also consider a weaker definition to identify issuers based on first six digits of the CUSIP and report the results in Column (23). The results also hold when I cluster by bond issue (Column (24)).

Next, I look to double cluster standard errors along two dimensions. If standard errors in yields are simultaneously correlated with state and advisor, I present results in Column (25) by double clustering errors. Likewise, I show results for clustering by advisor and issuer (Column (26)). Further, I shows robustness of the results to alternative specification involving double clustered standard errors along the cross-section and over time. Specifically, I double cluster by state and year in Column (27), and by advisor and year in Column (28). Finally, I modify the time dimension of clustering to year-month and report the results based on double clustering by state and year-month in Column (29), followed by advisor and year-month in Column (30). In all these specifications, the baseline result holds at the conventional levels of statistical significance, suggesting that the findings are robust to these alternative considerations of clustering standard errors.

Overall, I perform several robustness checks for the baseline specification and find that the main result is not driven by these alternative considerations/explanations. I show that the effect holds even after additional fixed effects, or stricter requirements involving smaller sub-samples. Moreover, I find similar results by using alternative approaches to clustering standard errors. These evidence enhance the argument of a causal interpretation of the main result in the paper. 

\clearpage

\clearpage
    \newpage

\begin{figure}
\begin{center}
\textbf{Total Issuance: Negotiated vs Competitive}
\end{center}
\centering
\includegraphics[scale=0.85]{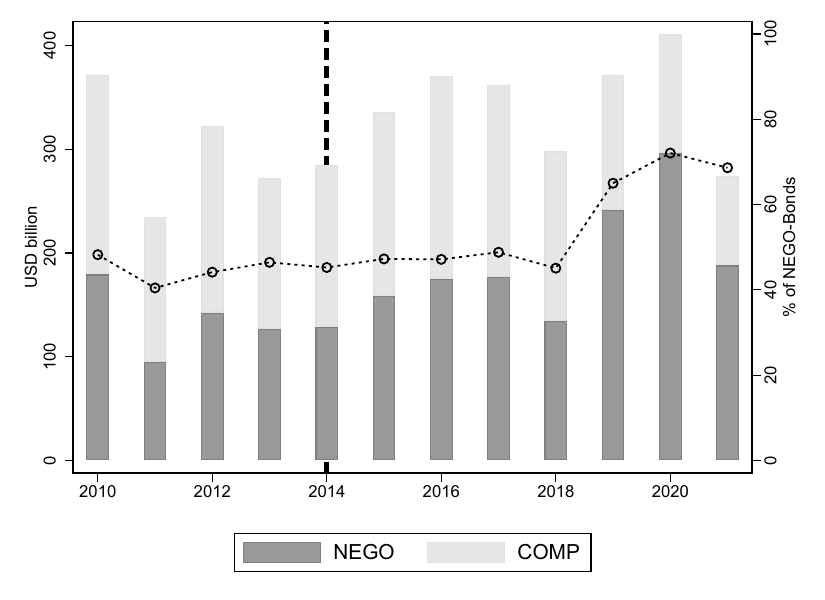}

\caption{\footnotesize{This figure shows the annual amount of advised bonds sold via negotiation. The left hand axis corresponds to the vertical bars denoting the amounts in USD billion. The line graph represents the percentage of dollar value sold via negotiation, shown on the right axis.}}
\label{fig:amt_nego}
\end{figure}

\begin{figure}
\begin{center}
\textbf{Likelihood of Negotiated Sale}
\end{center}
\centering
\includegraphics[scale=0.85]{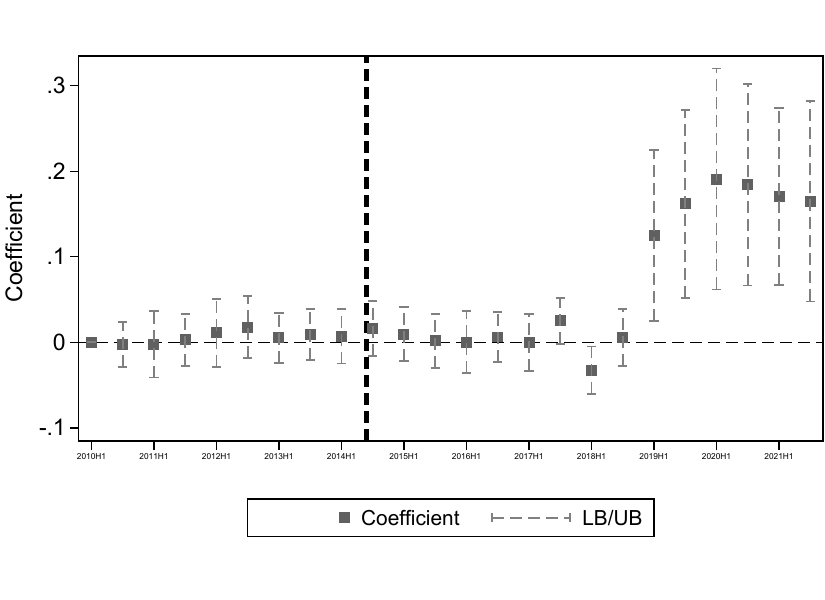}

\caption{\footnotesize{This figure reports the likelihood of issuing a bond via negotiation, \textit{within} issuer. The coefficients are shown relative to a benchmark period of half year period at the start of the event window. Standard errors are clustered at the state level.}}
\label{fig:ll_nego}
\end{figure}

\begin{landscape}
\begin{figure}
\begin{center}
\textbf{Municipal Advisor and Underwriter Network (2010-2021)}
\end{center}
\centering
\includegraphics[scale=0.60]{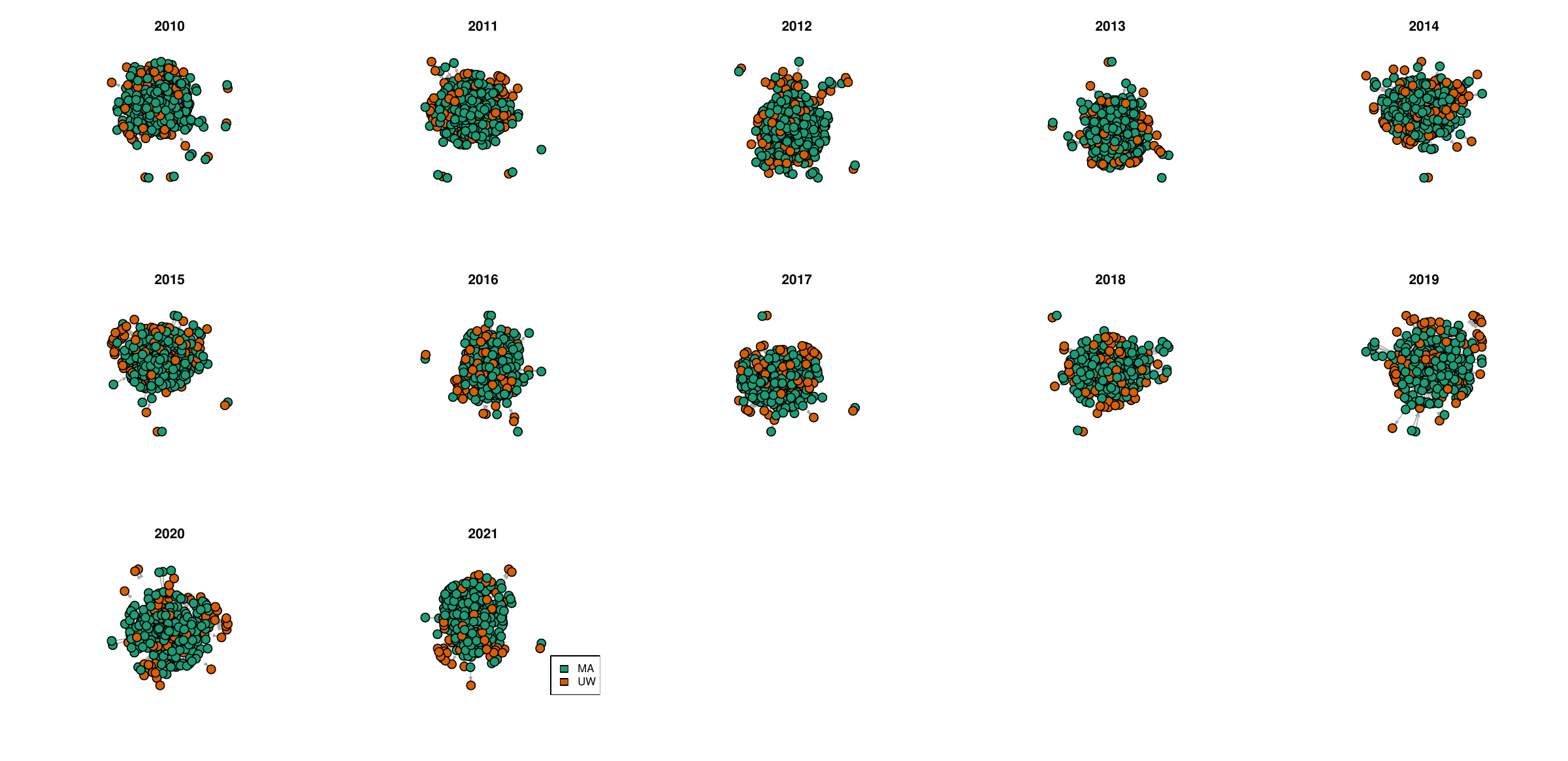}
\label{fig:network}

\caption{\footnotesize{The figure shows connections among municipal advisors and lead underwriters for each year during 2010-2021. Municipal advisors (MA) are denoted with green circles and underwriters (UW) with orange circles. Vertices are weighted by the fraction of par value issued by that MA-UW pair relative to the total par value issued by the MA in that year.}}
\end{figure}
\end{landscape}

\newpage
\clearpage

\begin{table}[htbp]
\def\sym#1{\ifmmode^{#1}\else\(^{#1}\)\fi}
\caption{Summary Statistics: Municipal Bonds}
\label{table:summary_bonds}
\vspace{0.2in}
\par
\footnotesize{This table summarizes the municipal bond level characteristics during 2010-2021 for the bonds in the sample. The two panels correspond to the treated (negotiated sale) versus control (competitively bid) bonds. The key variables are described in Table \ref{table:vardescription}.}\vspace{.2in} \\

\centering
\vspace{0.1in}
\resizebox{1.0\textwidth}{!} {

\begin{tabular}{l*{1}{cccccc}}
\hline
                    &\multicolumn{6}{c}{}                               \\
                    &       Count&        Mean&       Std. Dev.&   P25&  P50& P75         \\
\hline
\textbf{Treated - Negotiated Sale}              &            &            &            &            \\
Amount (USD million)&     319,128&        4.13&       16.30&         0.4&         1.0&         2.8\\
Coupon(\%)          &     319,128&        3.88&        1.09&         3.0&         4.0&         5.0\\
Years to Maturity   &     319,128&       10.54&        6.46&         5.5&         9.4&        14.6\\
Offering Price (USD)&     319,128&      109.74&        9.05&       101.6&       108.8&       116.4\\
Offering Yield(\%)  &     319,128&        2.33&        1.03&         1.5&         2.2&         3.0\\
Yield Spread(\%)    &     319,128&        1.23&        1.26&         0.4&         1.1&         2.0\\
Callable (Dummy)    &     319,128&        0.49&        0.50&         0.0&         0.0&         1.0\\
General Obligation (Dummy)&     319,128&        0.43&        0.49&         0.0&         0.0&         1.0\\
Bank Qualified (Dummy)&     319,128&        0.24&        0.42&         0.0&         0.0&         0.0\\
Cred. Enh. (Dummy)  &     319,128&        0.14&        0.35&         0.0&         0.0&         0.0\\
Insured (Dummy)     &     319,128&        0.21&        0.41&         0.0&         0.0&         0.0\\

\hline\\
\textbf{Control - Competitively Bid}              &            &            &            &            \\
Amount (USD million)&     618,866&        1.99&        8.15&         0.2&         0.5&         1.3\\
Coupon(\%)          &     618,866&        3.28&        1.12&         2.2&         3.0&         4.0\\
Years to Maturity   &     618,866&        9.92&        5.93&         5.0&         9.0&        13.9\\
Offering Price (USD)&     618,865&      105.95&        7.64&       100.0&       103.1&       109.6\\
Offering Yield(\%)  &     618,866&        2.21&        0.97&         1.5&         2.2&         2.9\\
Yield Spread(\%)    &     618,866&        1.22&        1.12&         0.4&         1.1&         2.0\\
Callable (Dummy)    &     618,866&        0.49&        0.50&         0.0&         0.0&         1.0\\
General Obligation (Dummy)&     618,866&        0.68&        0.47&         0.0&         1.0&         1.0\\
Bank Qualified (Dummy)&     618,866&        0.48&        0.50&         0.0&         0.0&         1.0\\
Cred. Enh. (Dummy)  &     618,866&        0.25&        0.43&         0.0&         0.0&         0.0\\
Insured (Dummy)     &     618,866&        0.16&        0.36&         0.0&         0.0&         0.0\\
\hline
\end{tabular}
}
\end{table}

\begin{table}[htbp]
\def\sym#1{\ifmmode^{#1}\else\(^{#1}\)\fi}
\caption{Impact on Offering Yield Spreads of Local Governments} 
\label{table:baseline_yld}
\vspace{0.2in}
\par
{\footnotesize This table reports the baseline results for the sample using Equation \eqref{eq:baseline} estimating the differential effect on municipal bond offering yields of treated and control bonds after the Municipal Advisor Rule of 2014. The primary coefficient of interest, $\beta_0$, is captured by the interaction term of \textit{Treated $\times$ Post}. I show the results using offering yield as the dependent variable. I provide the description of key variables in Table \ref{table:vardescription}. T-statistics are reported in brackets and standard errors are clustered at the state level. \sym{*} \(p<0.10\), \sym{**} \(p<0.05\), \sym{***} \(p<0.01\)}\\
\vspace{.05in} 

\vspace{0.1in}
\centering
\resizebox{1.0\textwidth}{!} {
\begin{tabular}{@{\extracolsep{4pt}}l*{5}{c}}
\hline
                    \emph{Dependent Variable}:&\multicolumn{5}{c}{Offering Yield (bps)}\\
                    \cline{2-6}\\                    &\multicolumn{1}{c}{(1)}&\multicolumn{1}{c}{(2)}&\multicolumn{1}{c}{(3)}&\multicolumn{1}{c}{(4)}&\multicolumn{1}{c}{(5)}\\
\hline
Treated $\times$ Post &      -10.58\sym{**} &      -10.58\sym{**} &      -13.78\sym{***}&       -9.99\sym{**} &      -10.26\sym{***}\\
                    &     [-2.51]         &     [-2.51]         &     [-3.52]         &     [-2.64]         &     [-2.92]         \\
[1em]
Treated             &       -0.97         &       -0.97         &       10.65\sym{***}&        8.60\sym{***}&        9.32\sym{***}\\
                    &     [-0.18]         &     [-0.18]         &      [3.46]         &      [2.91]         &      [3.32]         \\
[1em]
Post                &      -14.77\sym{***}&      -14.77\sym{***}&      -10.82\sym{***}&      -11.91\sym{***}&       -3.16         \\
                    &     [-7.00]         &     [-7.00]         &     [-4.87]         &     [-5.83]         &     [-1.63]         \\
[1em]
UW Mkt. Share(\%)   &                     &                     &                     &                     &      -13.16\sym{***}\\
                    &                     &                     &                     &                     &     [-3.05]         \\
[1em]
$\triangle$Fed Funds$_{t-1}$&                     &                     &                     &                     &      -16.93\sym{***}\\
                    &                     &                     &                     &                     &    [-13.95]         \\
[1em]
$\triangle$UST10Y$_{t-1}$&                     &                     &                     &                     &       26.01\sym{***}\\
                    &                     &                     &                     &                     &     [27.23]         \\
[1em]
\hline
Issuer FE           &$\checkmark$         &$\checkmark$         &$\checkmark$         &$\checkmark$         &$\checkmark$         \\
State-Yr. FE        &$\checkmark$         &$\checkmark$         &$\checkmark$         &$\checkmark$         &$\checkmark$         \\
Bond Controls       &                     &                     &$\checkmark$         &$\checkmark$         &$\checkmark$         \\
Rating FE       &                     &                     &$\checkmark$         &$\checkmark$         &$\checkmark$         \\
County Controls     &                     &                     &$\checkmark$         &$\checkmark$         &$\checkmark$         \\
Adviser-Yr. FE      &                     &                     &                     &$\checkmark$         &$\checkmark$         \\
Adj.-R$^2$          &       0.323         &       0.323         &       0.857         &       0.865         &       0.872         \\
Obs.                &     937,845         &     937,845         &     937,845         &     937,819         &     937,819         \\
\hline
\end{tabular}
}
\end{table}

\clearpage
\newpage
\begin{table}[htbp]
\def\sym#1{\ifmmode^{#1}\else\(^{#1}\)\fi}
\caption{Robustness Tests}
\label{table:robustness}
\vspace{0.2in}
\par
{\footnotesize In this table I report results for various robustness tests on the baseline specification, i.e., Column (5) of Table \ref{table:main_baseline}. In Columns (1)-(9), I report results based on alternative econometric specifications. I introduce additional fixed effects to account for unobserved factors that may be varying over time. Specifically, Column (1) reports baseline results by adding issuer-type $\times$ year fixed effects. Column (2) shows results by adding bond purpose fixed effects. I add bond purpose $\times$ year fixed effect in Column (3). Columns (4) and (5) show results by adding underwriter fixed effect and underwriter $\times$ year fixed effect, respectively. In Columns (6) and (7), I control for unobserved pairing between issuers and advisors, as well as issuers and underwriters, separately, by adding issuer-advisor pair fixed effect and issuer-underwriter pair fixed effect, respectively. These specifications include time unvarying advisor and underwriter fixed effects, respectively. Column (8) shows results with advisor $\times$ state fixed effects. In Column (9), I also add county $\times$ year fixed effects. I consider additional tax considerations in Columns (10)-(12). First, I relax the sample of bonds to include taxable bonds in Columns (10)-(11). In Column (11), I further omit bonds from five states (IL, IA, KS, OK, WI) that tax interest income on municipal bonds issued in-state or out-of-state \citep{gao2021good}. Column (12) shows results using bonds that are exempt from both state and federal income tax simultaneously. Columns (13)-(16) report results focusing on sub-samples of homogeneous bonds. Accordingly, in Column (13), I drop bonds in which the advisor and underwriter are same. Column (14) shows results by dropping callable bonds. I drop insured bonds in Column (15). Finally, I focus on only the new money bonds in Column (16). The results in Columns (17)-(19) focus on additional geographic considerations. I keep only local bonds (by dropping state level bonds) in Column (17). Conversely, I show results using only the state level bonds in Column (18). In Column (19), I report the baseline results by dropping issuances from the three largest municipal bond issuers, namely: California (CA), New York (NY) and Texas (TX). I consider alternative levels of clustering standard errors in Columns (20)-(30). In Columns (20)-(24), I cluster standard errors by advisor, underwriter, issuer, issuer(2), and bond issue, respectively. Issuer(2) in Column (23) refers to weakly identifying borrowers based on the first six-digits of the bond CUSIP \citep{gao2021good}. Columns (25)-(26) double cluster standard errors along two dimensions in the geography of issuers: state-advisor, and advisor-issuer, respectively. Finally, in Columns (27)-(30), I double cluster standard errors over time and across bonds using: state and year, advisor and year, state and year-month, and advisor and year-month, respectively. T-statistics are reported in brackets and standard errors are clustered at the state level, unless otherwise specified. \sym{*} \(p<0.10\), \sym{**} \(p<0.05\), \sym{***} \(p<0.01\)}\\
\end{table}
\clearpage
\newpage
\begin{landscape}
\begin{table}
    \centering
    \def\sym#1{\ifmmode^{#1}\else\(^{#1}\)\fi}
    \vspace{0.2cm}
    \resizebox{1.3\textwidth}{!}{
\begin{tabular}{@{\extracolsep{4pt}}l*{9}{c}@{}}
\hline
                    \emph{Dependent Variable}: &\multicolumn{9}{c}{ Yield Spread (basis points)} \vspace{0.2in} \\
                    &\multicolumn{9}{c}{Alternative Specifications}\\
                    \cline{2-10}  \\
                    &\multicolumn{1}{c}{Add Issuer-Type}&\multicolumn{1}{c}{Add}&\multicolumn{1}{c}{Add Purpose}&\multicolumn{1}{c}{Add }&\multicolumn{1}{c}{Add }&\multicolumn{1}{c}{Add }&\multicolumn{1}{c}{Add}&\multicolumn{1}{c}{Add}&\multicolumn{1}{c}{Add}\\           &\multicolumn{1}{c}{Year FE}&\multicolumn{1}{c}{Purpose FE}&\multicolumn{1}{c}{Year FE}&\multicolumn{1}{c}{UW FE}&\multicolumn{1}{c}{UW-Yr. FE}&\multicolumn{1}{c}{Iss.-MA FE}&\multicolumn{1}{c}{Iss.-UW FE}&\multicolumn{1}{c}{MA-State FE}&\multicolumn{1}{c}{County-Yr. FE}\\ &\multicolumn{1}{c}{(1)}&\multicolumn{1}{c}{(2)}&\multicolumn{1}{c}{(3)}&\multicolumn{1}{c}{(4)}&\multicolumn{1}{c}{(5)}&\multicolumn{1}{c}{(6)}&\multicolumn{1}{c}{(7)}&\multicolumn{1}{c}{(8)}&\multicolumn{1}{c}{(9)}\\
\hline\\
\vspace{0.2cm}
Treated $\times$ Post &       -9.99\sym{***}&      -10.61\sym{***}&      -10.08\sym{***}&      -11.15\sym{***}&       -9.26\sym{***}&      -12.27\sym{***}&      -14.02\sym{***}&       -9.98\sym{***}&       -8.79\sym{**} \\
                    &     [-3.09]         &     [-3.29]         &     [-3.00]         &     [-3.82]         &     [-3.24]         &     [-3.42]         &     [-2.79]         &     [-3.08]         &     [-2.16]         \\
[1em]
Adj.-R$^2$          &       0.850         &       0.850         &       0.850         &       0.851         &       0.854         &       0.853         &       0.884         &       0.851         &       0.867         \\
Obs.                &     936,666         &     935,293         &     935,293         &     937,813         &     937,798         &     937,715         &     937,291         &     937,805         &     937,684         \\
\hline
\end{tabular}
}\\

\medskip
   \resizebox{1.3\textwidth}{!}{
\begin{tabular}{@{\extracolsep{4pt}}l*{10}{c}@{}}

                    \emph{Dependent Variable}: &\multicolumn{10}{c}{ Yield Spread (basis points)} \vspace{0.2in}\\
                    \cline{2-11} \\
                    &\multicolumn{3}{c}{Tax Considerations}&\multicolumn{4}{c}{Bond Considerations}&\multicolumn{3}{c}{Geographic Considerations}\\
                    \cline{2-4}  \cline{5-8} \cline{9-11} \\ &\multicolumn{2}{c}{Include Taxable}&\multicolumn{1}{c}{Exempt}&\multicolumn{1}{c}{Drop same}&\multicolumn{1}{c}{Drop}&\multicolumn{1}{c}{Drop}&\multicolumn{1}{c}{Only}&\multicolumn{1}{c}{Keep}&\multicolumn{1}{c}{Keep}&\multicolumn{1}{c}{Drop}\\     \cline{2-3}\vspace{-1em}\\  &\multicolumn{1}{c}{}&\multicolumn{1}{c}{(-Taxable States)}&\multicolumn{1}{c}{(Fed.+State)}&\multicolumn{1}{c}{Adv.-UW. bonds}&\multicolumn{1}{c}{Callable}&\multicolumn{1}{c}{Insured}&\multicolumn{1}{c}{New money}&\multicolumn{1}{c}{Local}&\multicolumn{1}{c}{State}&\multicolumn{1}{c}{CA,NY,TX}\\                    &\multicolumn{1}{c}{(10)}&\multicolumn{1}{c}{(11)}&\multicolumn{1}{c}{(12)}&\multicolumn{1}{c}{(13)}&\multicolumn{1}{c}{(14)}&\multicolumn{1}{c}{(15)}&\multicolumn{1}{c}{(16)}&\multicolumn{1}{c}{(17)}&\multicolumn{1}{c}{(18)}&\multicolumn{1}{c}{(19)}\\
\hline\\
\vspace{0.2cm}
Treated $\times$ Post &      -10.62\sym{***}&       -9.27\sym{***}&      -10.27\sym{***}&      -10.91\sym{***}&      -10.29\sym{***}&      -13.28\sym{***}&      -21.99\sym{***}&       -9.29\sym{***}&      -12.85\sym{*}  &      -15.00\sym{***}\\
                    &     [-3.33]         &     [-3.01]         &     [-3.16]         &     [-3.62]         &     [-4.59]         &     [-4.88]         &     [-4.72]         &     [-2.83]         &     [-1.89]         &     [-5.46]         \\
[1em]
Adj.-R$^2$          &       0.850         &       0.849         &       0.849         &       0.849         &       0.873         &       0.853         &       0.873         &       0.856         &       0.838         &       0.853         \\
Obs.                &     937,819         &     828,142         &     864,320         &     933,485         &     479,225         &     775,293         &     473,494         &     845,881         &      91,928         &     585,319         \\
\hline
\end{tabular}
}\\

\medskip
   \resizebox{1.3\textwidth}{!}{
\begin{tabular}{@{\extracolsep{4pt}}l*{11}{c}@{}}
                    \emph{Dependent Variable}: &\multicolumn{11}{c}{ Yield Spread (basis points)} \vspace{0.2in} \\
                    \cline{2-12} \\
                    &\multicolumn{11}{c}{Alternative Clustering}\\
                    \cline{2-12}\\                    &\multicolumn{1}{c}{Advisor}&\multicolumn{1}{c}{Underwriter}&\multicolumn{1}{c}{Issuer}&\multicolumn{1}{c}{Issuer(2)}&\multicolumn{1}{c}{Issue}&\multicolumn{1}{c}{State, Advisor}&\multicolumn{1}{c}{Advisor, Issuer}&\multicolumn{1}{c}{State, Year}&\multicolumn{1}{c}{Advisor, Year}&\multicolumn{1}{c}{State, YM}&\multicolumn{1}{c}{Advisor, YM}\\    &\multicolumn{1}{c}{(20)}&\multicolumn{1}{c}{(21)}&\multicolumn{1}{c}{(22)}&\multicolumn{1}{c}{(23)}&\multicolumn{1}{c}{(24)}&\multicolumn{1}{c}{(25)}&\multicolumn{1}{c}{(26)}&\multicolumn{1}{c}{(27)}&\multicolumn{1}{c}{(28)}&\multicolumn{1}{c}{(29)}&\multicolumn{1}{c}{(30)}\\
\hline\\
\vspace{0.2cm}
Treated $\times$ Post &      -10.62\sym{***}&      -10.62\sym{***}&      -10.62\sym{***}&      -10.62\sym{***}&      -10.62\sym{***}&      -10.62\sym{***}&      -10.62\sym{***}&      -10.62\sym{**} &      -10.62\sym{***}&      -10.62\sym{**} &      -10.62\sym{***}\\
                    &     [-4.13]         &     [-5.71]         &     [-5.80]         &     [-7.85]         &     [-9.33]         &     [-3.10]         &     [-3.80]         &     [-2.92]         &     [-3.37]         &     [-2.68]         &     [-3.91]         \\
[1em]
Adj.-R$^2$          &       0.850         &       0.850         &       0.850         &       0.850         &       0.850         &       0.850         &       0.850         &       0.850         &       0.850         &       0.850         &       0.850         \\
Obs.                &     937,819         &     937,819         &     937,819         &     937,819         &     937,819         &     937,819         &     937,819         &     937,819         &     937,819         &     937,819         &     937,819         \\
\hline
\end{tabular}
}
\end{table}
\end{landscape}

\begin{landscape}
\begin{table}[htbp]
\def\sym#1{\ifmmode^{#1}\else\(^{#1}\)\fi}
\caption{Examples Showing Advisor's Ex-ante Role in ``Selecting'' Underwriter} 
\label{table:eg_byUWintro}
\vspace{0.2in}
\par
{\footnotesize This table shows some examples of advisor's ex-ante role in ``selecting'' underwriters. The first column shows the name of issuers, followed by their state in the second column. ``Adv.'' refers to the indicator for the use of municipal advisor in the bond issuance. The fourth column (``Intro.'') shows a dummy indicating bond issues where the municipal advisor introduces a new underwriter. The subsequent columns indicate the name of the municipal advisor, the underwriter, and the offering date of the municipal bond issue, respectively.}\\
\vspace{.05in} 

\vspace{0.1in}
\centering
\resizebox{1.3\textwidth}{!} {
    \begin{tabular}{llrrllr}
    \textbf{Issuer} & \textbf{State} & \multicolumn{1}{l}{\textbf{Adv.}} & \multicolumn{1}{l}{\textbf{Intro.}} & \textbf{Municipal Advisor} & \textbf{Underwriter} & \multicolumn{1}{l}{\textbf{Offering Date}} \\
    \hline
    Anaheim Union High School District & CA    & 0     & 0     &       & BANC OF AMERICA  & 27-Aug-04 \\
              &       &       &       &       &       &  \\
Anaheim Union High School District & CA    & 1     & 0     & CALDWELL FLORES WINTERS INC & BANC OF AMERICA & 27-Aug-04 \\
          &       &       &       &       &       &  \\
    Anaheim Union High School District & CA    & 1     & 1     & CALDWELL FLORES WINTERS INC & UBS FINANCIAL & 14-Jan-05 \\
              &       &       &       &       &       &  \\
Anaheim Union High School District & CA    & 1     & 0     & CALDWELL FLORES WINTERS INC & UBS FINANCIAL  & 10-May-06 \\
          &       &       &       &       &       &  \\
    Anaheim Union High School District & CA    & 1     & 1     & GOVT. FINANCIAL STRATEGIES INC & JP MORGAN CHASE  & 11-Oct-12 \\
          &       &       &       &       &       &  \\
          &       &       &       &       &       &  \\
          &       &       &       &       &       &  \\          
    Allegheny County Airport Authority & PA    & 0     & 0     &       & BANC OF AMERICA  & 30-Nov-06 \\
              &       &       &       &       &       &  \\
Allegheny County Airport Authority & PA    & 0     & 0     &       & BANC OF AMERICA  & 11-Sep-07 \\
          &       &       &       &       &       &  \\
    Allegheny County Airport Authority & PA    & 1     & 1     & MORGAN KEEGAN & PIPER JAFFRAY \& CO. & 12-Aug-10 \\
          &       &       &       &       &       &  \\
    Allegheny County Airport Authority & PA    & 1     & 1     & RAYMOND JAMES & JEFFERIES GROUP LLC & 24-Apr-12 \\
          &       &       &       &       &       &  \\
          &       &       &       &       &       &  \\
          &       &       &       &       &       &  \\          
    Alvord Unified School District & CA    & 0     & 0     &       & STONE \& YOUNGBERG & 15-Mar-07 \\
              &       &       &       &       &       &  \\
Alvord Unified School District & CA    & 1     & 1     & DOLINKA GROUP INC & UBS FINANCIAL & 17-Apr-08 \\
          &       &       &       &       &       &  \\
    Alvord Unified School District & CA    & 1     & 1     & DOLINKA GROUP INC & PIPER JAFFRAY \& CO. & 1-Jun-09 \\
              &       &       &       &       &       &  \\
Alvord Unified School District & CA    & 1     & 0     & DOLINKA GROUP INC & PIPER JAFFRAY \& CO. & 13-May-10 \\
          &       &       &       &       &       &  \\
    Alvord Unified School District & CA    & 1     & 0     & DOLINKA GROUP INC & PIPER JAFFRAY \& CO. & 1-Jun-11 \\
    \hline
    \end{tabular}%
}
\end{table}
\end{landscape}

\end{appendices}

\end{document}